\documentclass[twocolumn,english,aps,prl,superscriptaddress]{revtex4-1}
\usepackage[latin9]{inputenc}
\usepackage{color}
\usepackage{mathrsfs}
\usepackage{amsmath}
\usepackage{graphicx}
\usepackage{times}
\usepackage{mathptmx}

\makeatletter
\usepackage[colorlinks,citecolor=blue,linkcolor=blue]{hyperref}

\makeatother

\usepackage{babel}
\begin{document}
\title{Microscopic origin of molecule excitation via inelastic electron scattering
in scanning tunneling microscope}
\author{Guohui Dong}
\affiliation{Graduate School of Chinese Academy of Engineering Physics, Beijing
100084, China}
\author{Yining You}
\affiliation{Department of Modern Physics, University of Science and Technology
of China, Hefei 230026, China}
\author{Hui Dong}
\email{hdong@gscaep.ac.cn}

\affiliation{Graduate School of Chinese Academy of Engineering Physics, Beijing
100084, China}
\begin{abstract}
The scanning-tunneling-microscope-induced luminescence emerges recently
as an incisive tool to measure the molecular properties down to the
single-molecule level. The rapid experimental progress is far ahead
of the theoretical effort to understand the observed phenomena. Such
incompetence leads to a significant difficulty in quantitatively assigning
the observed feature of the fluorescence spectrum to the structure
and dynamics of a single molecule. This letter is devoted to reveal
the microscopic origin of the molecular excitation via inelastic scattering
of the tunneling electrons in scanning tunneling microscope. The current
theory explains the observed large photon counting asymmetry between
the molecular luminescence intensity at positive and negative bias
voltage.
\end{abstract}
\maketitle
\textit{Introduction} -- The physical limitation of conventional
semiconductor devices spurs the recent development of single molecule
photoelectronics \citep{Aradhya_2013,XinNa_2019,sunlanlan_single_molecule_2014},
where the incisive tool to probe single molecular structure and dynamics
is of great demand. Combining the high resolution of scattering tunneling
microscope (STM) with the specificity of fluorescence spectroscopy
of molecules, STM-induced luminescence (STML) provides an ideal tool
to study the photon emission and dynamics on the single-molecule level
\citep{Eli_science_1993,Berndt_science_1993}. Experimental breakthroughs
have allowed direct observations of the single-molecular properties,
e.g., the dipole-dipole coupling between molecules \citep{YangZhang_Nature_2016_dipole_dipole,Doppagne_PRL_2017,YangLuo_Single_photon_superradiance_PRL_2019},
the energy transfer in molecular dimers \citep{Imada_nature_2016_energy_transfer},
and the Fano-like lineshape \citep{HiroshiImada_fano_PRL_2017,YaoZhang_NC_fano_lineshape_2017,Kroger_nanolett_fano_2018}.
Yet,\textbf{ }\textcolor{black}{the retarded theoretical followup
prevents us from conclusively understanding the single-molecular properties
through} the quantitative analyses of experimental data\textcolor{black}{.}

Such lag of the corresponding theoretical effort has led to inconsistent
between experimental explanations. The underlying origin of the asymmetric
emission intensity at positive and negative bias between the tip and
substrate was assigned as the carrier-injection mechanism in \citep{YangZhang_Nature_2016_dipole_dipole},
while it was also understood as inelastic electron tunneling (probably
mediated by the localized surface plasmon) \citep{Doppagne_science_2018}
for the same molecule, i.e., the single ZnPc molecule. The question
exists even on the asymmetry with larger tunneling current at positive
bias or versa \citep{YangZhang_Nature_2016_dipole_dipole,Doppagne_science_2018}.
The inconsistency remains unresolved mainly due to the lack of microscopic
theory to conclusively determine the properties of the different tunneling
mechanisms, which are mixed in the \textit{ab initio} calculations
\citep{simulation_xyWu_2019,simulation_Miwa_2019}.

In this letter, we reveal the underlying microscopic origin of the
inelastic electron scattering down to the basic Coulomb interaction
between the tunneling electron and the single molecule. Our theory
shows the asymmetry with larger tunneling current and photon counting
rate at negative bias, in turn, excludes the possibility of the opposite
asymmetry to be attributed to the inelastic electron scattering. Such
attempt shall initiate the understanding of the experimental feature
from its microscopic origin and stimulate the theoretical studies
of the STML.

\begin{figure}
\begin{centering}
\includegraphics[scale=0.37]{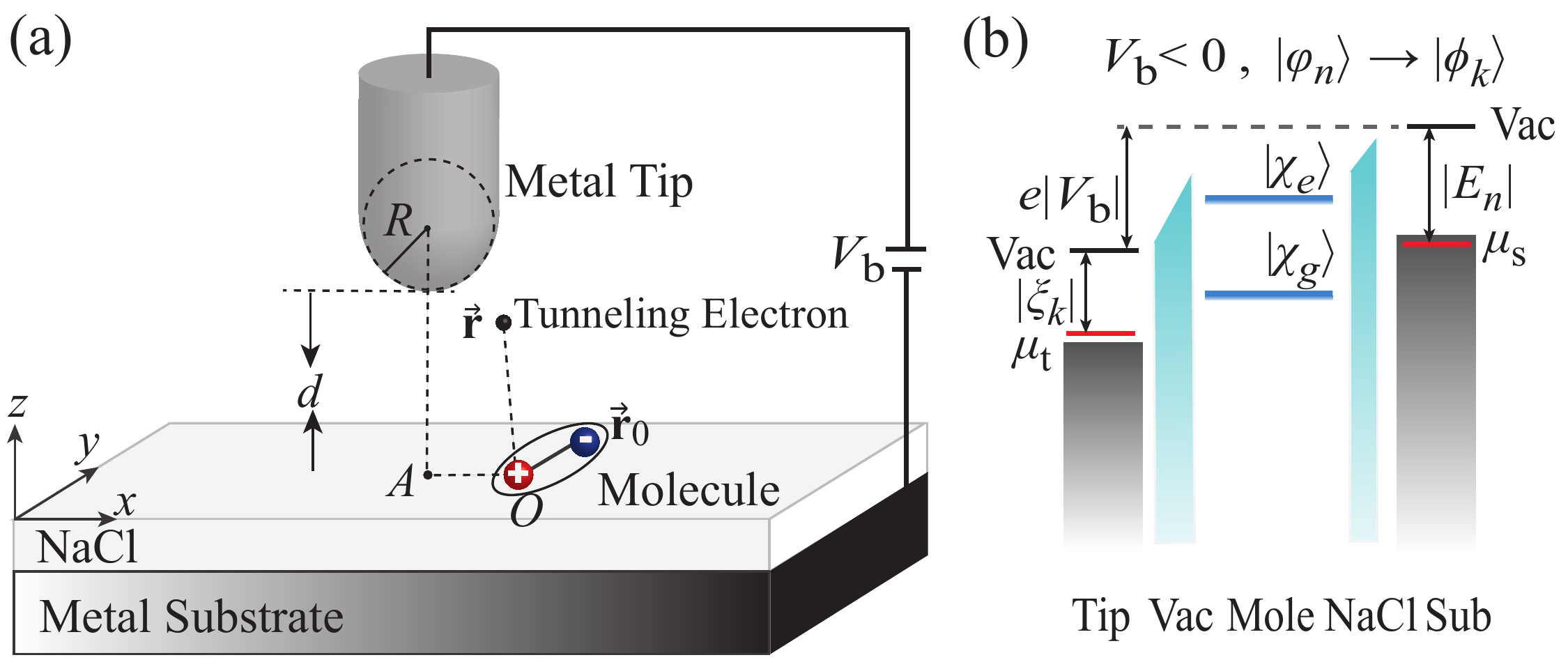}
\par\end{centering}
\caption{(Color online) (a) Schematic diagram of STML of a single molecule
placed on a salt-covered metal plane. The STM tip apex is modeled
as a sphere with radius $R$. Point $A$ is the projection of tip's
center on the plane, and $d$ is the distance between tip and plane.
The position of the positive charge in the molecule (red) is set as
the origin of the coordinate system. $\vec{\mathbf{r}}$ and $\vec{\mathbf{r}}_{0}$
stand for the vector of the tunneling electron (black) and the negative
charge in molecule (blue), respectively. (b) The level diagram for
inelastic electron scattering mechanism at negative bias. The black
lines denote the vacuum level at two electrodes, and the red lines
represent the initial and final electronic states. $\mu_{\mathrm{t}}\equiv\mu_{\mathrm{0}}+eV_{\mathrm{b}}$
and $\mu_{\mathrm{s}}\equiv\mu_{\mathrm{0}}$ are the Fermi energies
of tip and substrate at bias voltage $V_{\mathrm{b}}$, where $\mu_{0}$
is the Fermi energy of tip and substrate at zero bias.}

\label{fig:Schematic Diagram}
\end{figure}

\textit{Model} -- For the clarity of the notation, we sketch the
design of the single-molecule STML in Fig. \ref{fig:Schematic Diagram}(a).
A molecule, simplified for clarity as a dipole with positive (red)
and negative (blue) charge, is deposited on a salt-covered metal substrate.
A metal tip is positioned above the substrate plane. Both the tip
and substrate are typically used with noble meta, e.g., silver (Ag).
With nonzero bias voltage, an electron (black) from one electrode
excites the molecule via the Coulomb interaction during its tunneling
through the vacuum and then enters into the other electrode (see Fig.
\ref{fig:Schematic Diagram}(b)). Subsequently, the excited molecule
emits a photon by the spontaneous emission, which is measured by the
photon counting to reveal molecular properties.

The Hamiltonian for the setup is divided into three parts as $H=H_{\mathrm{el}}+H_{\mathrm{m}}+H_{\mathrm{el-m}}$,
where $H_{\mathrm{el}}$ is the Hamiltonian for the tunneling electron
between the tip and substrate, $H_{\mathrm{m}}$ is the Hamiltonian
of the molecule, and $H_{\mathrm{el-m}}$ is the interaction between
the tunneling electron and the single molecule. The Hamiltonian of
the tunneling electron is $H_{\mathrm{el}}=-\nabla^{2}/\left(2m_{e}\right)+V\left(\vec{\mathbf{r}}\right)$,
where $V\left(\vec{\mathbf{r}}\right)$ is the potential for the tunneling
electron at position $\vec{\mathbf{r}}=(x,y,z)$ and $m_{e}$ is the
mass of an electron. The wave functions are written for different
regions \citep{bardeen_1961,TUTORIAL_2006} as

\begin{subequations}
\begin{align}
H_{\mathrm{el,t}}\left|\phi_{k}\right\rangle  & \simeq\widetilde{\xi}_{k}\left|\phi_{k}\right\rangle ,\label{eq:solution_tip=000026base_nonzero_bias}\\
H_{\mathrm{el,s}}\left|\varphi_{n}\right\rangle  & \simeq\widetilde{E}_{n}\left|\varphi_{n}\right\rangle ,
\end{align}
\end{subequations}where $H_{\mathrm{el,t}}$ ($H_{\mathrm{el,s}}$)
is the Hamiltonian of the free tip (substrate) obtained by neglecting
the potential in the substrate (tip) region. $\left|\phi_{k}\right\rangle \left(\left|\varphi_{n}\right\rangle \right)$
is the eigenstate of free tip (substrate) with $\widetilde{\xi}_{k}\equiv\xi_{k}+eV_{\mathrm{b}}\left(\widetilde{E}_{n}\equiv E_{n}\right)$
where $\xi_{k}\left(E_{n}\right)$ is the eigenenergy with zero bias
voltage. The detailed form of the wave functions are discussed in
the supplementary material. The Hamiltonian for the molecule is simplified
as a two-level system \citep{Nian_simulation_2018,Nian_simu_2019}
$H_{\mathrm{m}}=E_{e}\left|\chi_{e}\right\rangle \left\langle \chi_{e}\right|+E_{g}\left|\chi_{g}\right\rangle \left\langle \chi_{g}\right|$,
where $|\chi_{e}\rangle\:(|\chi_{g}\rangle)$ is its excited (ground)
state with energy $E_{e}\left(E_{g}\right)$.

The key element to understand the mechanism is the interaction between
the molecule and the tunneling electron. For the purpose of clarity,
we consider a simple case of one tunneling electron. The interaction,
simplified from the Coulomb interaction, resembles the dipole interaction
as

\begin{align}
H_{\mathrm{el-m}} & \simeq-e\frac{\vec{\mathbf{\mu}}\cdot\vec{\mathbf{r}}}{\left|\vec{\mathbf{r}}\right|^{3}},\label{eq:electron_dipole_interaction}
\end{align}
where $\vec{\mathbf{\mu}}=-Ze\vec{\mathbf{r}}_{0}$ denotes the effective
electric dipole moment of the molecule. $Z$ is the effective charge
number, and $\vec{\mathbf{r}}_{0}$ stands for the vector of the center
of the electrons in molecule. $\vec{\mathbf{r}}$ represents the vector
of the tunneling electron. Here, we have chosen the central position
of the positive charge of molecule as the origin of the coordinate
system. The detailed derivation can be found for the molecule with
multiple chemical bonds \citep{Minkin_Dipole_Moment_1970} in the
supplementary material.

The interaction is rewritten explicitly with the basis of the wave
functions of the single molecule and tunneling electron as
\begin{align}
H_{\mathrm{el-m}} & =\sum_{n,k}\mathscr{N}_{\mathrm{s,t}}|_{V_{\mathrm{b}},E_{n}\rightarrow\xi_{k}}\sigma_{x}\left|\phi_{k}\right\rangle \left\langle \varphi_{n}\right|\nonumber \\
 & =\sum_{n,k}\mathscr{N}_{\mathrm{t,s}}|_{V_{\mathrm{b}},\xi_{k}\rightarrow E_{n}}\sigma_{x}\left|\varphi_{n}\right\rangle \left\langle \phi_{k}\right|.\label{eq:He-m}
\end{align}
We have defined the transition matrix element $\mathscr{N}_{\mathrm{s,t}}|_{V_{\mathrm{b}},E_{n}\rightarrow\xi_{k}}\equiv-e\vec{\mathbf{\mu}}\cdot\left\langle \phi_{k}\right|\vec{\mathbf{r}}/\left|\vec{\mathbf{r}}\right|^{3}\left|\varphi_{n}\right\rangle $
from substrate's state $\left|\varphi_{n}\right\rangle $ to tip's
state $\left|\phi_{k}\right\rangle $ and $\mathcal{N}_{\mathrm{t,s}}|_{V_{\mathrm{b}},\xi_{k}\rightarrow E_{n}}\equiv-e\vec{\mathbf{\mu}}\cdot\left\langle \varphi_{n}\right|\vec{\mathbf{r}}/\left|\vec{\mathbf{r}}\right|^{3}\left|\phi_{k}\right\rangle $
from tip's state $\left|\phi_{k}\right\rangle $ to substrate's state
$\left|\varphi_{n}\right\rangle $. $\sigma_{x}\equiv\left|\chi_{e}\right\rangle \left\langle \chi_{g}\right|+\left|\chi_{g}\right\rangle \left\langle \chi_{e}\right|$
is the transition matrix between molecular ground and excited states.
The electron-dipole interaction in Eq. (\ref{eq:He-m}) will induce
energy transfer between the tunneling electron and the molecule (the
state of the two-level molecule is flipped).

Tip's wave function in the vacuum region has the asymptotic spherical
form $\phi_{k}\left(\vec{\mathbf{r}}\right)=A_{k}e^{-\kappa_{k}\left|\vec{\mathbf{r}}-\vec{\mathbf{a}}\right|}/\left(\kappa_{k}\left|\vec{\mathbf{r}}-\vec{\mathbf{a}}\right|\right)$
where $\vec{\mathbf{a}}$ is the position of tip's center of curvature
and $\kappa_{k}=\sqrt{-2m_{e}\xi_{k}}$ is its decay factor. The normalized
coefficient $A_{k}$ can be determined by first-principles calculations.
This wave function is typical known as the s-wave, which is the simplest
case for the tip \citep{bardeen_1961,chen_1990}. Contribution from
other wave functions can be similarly considered as that in the studies
of STM \citep{chen_1990}. And substrate's wave function $\varphi_{n}\left(\vec{\mathbf{r}}\right)=B_{n}e^{-\kappa_{n}\left|z\right|}$
decays along the $+z$ direction with decay factor $\kappa_{n}=\sqrt{-2m_{e}E_{n}}$
\citep{Tersoff_Hamann_1983,Tersoff_Hamann_1985} and the normalization
constant $B_{n}$. With the wave functions for the tip and substrate,
the transition matrix element is explicitly written as

\begin{widetext}
\begin{align}
\mathscr{N}_{\mathrm{s,t}}|_{V_{\mathrm{b}},E_{n}\rightarrow\xi_{k}} & \simeq-A_{k}B_{n}\sum_{l=x,y,z}e\mu_{l}\int_{-\infty}^{\infty}dx\int_{-\infty}^{\infty}dy\int_{0}^{d}dzl\frac{e^{-\kappa_{n}z-\kappa_{k}\sqrt{(x-a_{x})^{2}+y^{2}+(z-d-R)^{2}}}}{(x^{2}+y^{2}+z^{2})^{3/2}},
\end{align}

\end{widetext}where $\mu_{x(y,z)}$ is the $x(y,z)$ component of
the molecular dipole moment. And \textcolor{black}{without loss of
generality}, we have chosen the position of tip's center of curvature
along $x$ axis, i.e., $\vec{\mathbf{a}}=\left(a_{x},0,d+R\right)$.
By taking the decay wave functions of tip and substrate into account,
we integrate over the region between plane $z=0$ and $z=d$ as an
approximation. And in the later discussion, we ignore the dependence
of $\mathscr{N}_{\mathrm{s,t}}|_{V_{\mathrm{b}},E_{n}\rightarrow\xi_{k}}$
on the normalization constants $A_{k}$ and $B_{n}$ by taking them
to independent on the index $k$ and $n$.

\textit{Asymmetry of photon counting} -- To understand the asymmetry
of photon counting, we calculate the tunneling rate at negative bias
($V_{\mathrm{b}}<0$), illustrated in Fig. \ref{fig:Schematic Diagram}(b),
where the Fermi level of tip is lower than that of substrate. The
molecule is initially in its ground state and the tunneling electron
in one of substrate's eigenstate, i.e., $\left|\Psi\left(t=0\right)\right\rangle =\left|\chi_{g}\right\rangle \left|\varphi_{n}\right\rangle $.
To the first order of $H_{\mathrm{el}}-H_{\mathrm{el,s}}$ and $H_{\mathrm{el-m}}$,
we obtain the time evolution of the system as
\begin{align}
\left|\Psi\left(t\right)\right\rangle  & =e^{-i\left(\widetilde{E}_{n}+E_{g}\right)t}\left|\chi_{g}\right\rangle \left|\varphi_{n}\right\rangle \nonumber \\
 & +\sum_{k}c_{g,k}\left(t\right)\left|\chi_{g}\right\rangle \left|\phi_{k}\right\rangle +\sum_{k}c_{e,k}\left(t\right)\left|\chi_{e}\right\rangle \left|\phi_{k}\right\rangle ,\label{eq:state_at_time_t}
\end{align}
where the second and third terms stand for elastic and inelastic tunneling
respectively. In order to obtain the above result, we have applied
the rotating-wave approximation for Hamiltonian in Eq. (\ref{eq:He-m}).
The corresponding tunneling amplitudes read

\begin{align}
c_{g,k}\left(t\right) & =e^{-iE_{g}t}\frac{e^{-i\widetilde{E}_{n}t}-e^{-i\widetilde{\xi}_{k}t}}{\widetilde{E}_{n}-\widetilde{\xi}_{k}}\mathscr{M}_{n,k},\label{eq:cgk}\\
c_{e,k}\left(t\right) & =\frac{e^{-i\left(\widetilde{E}_{n}+E_{g}\right)t}-e^{-i\left(\widetilde{\xi}_{k}+E_{e}\right)t}}{\widetilde{E}_{n}-\widetilde{\xi}_{k}-E_{eg}}\mathscr{N}_{\mathrm{s,t}}|_{V_{\mathrm{b}},E_{n}\rightarrow\xi_{k}},\label{eq:cek}
\end{align}
where $\mathscr{M}_{n,k}\equiv\left\langle \phi_{k}\right|\left(H_{\mathrm{el}}-H_{\mathrm{el,s}}\right)\left|\varphi_{n}\right\rangle $
is the transition matrix element of the elastic tunneling and $E_{eg}\equiv E_{e}-E_{g}$
is the optical gap of the single molecule.

We will focus on the inelastic tunneling process instead of the elastic
tunneling which has been well explored in the earlier development
\citep{bardeen_1961,Tersoff_Hamann_1983,Tersoff_Hamann_1985,chen_1990}
of STM. The inelastic tunneling rate $\mathscr{J}_{n\rightarrow k}$
from $\left|\varphi_{n}\right\rangle $ to $\left|\phi_{k}\right\rangle $
is $\mathscr{J}_{n\rightarrow k}=d\left|c_{e,k}\left(t\right)\right|^{2}/dt$.
The overall inelastic electron current at negative voltage $I_{-\mathrm{,inela}}=e\sum_{n}\sum_{k}\mathscr{J}_{n\rightarrow k}F_{\mu_{\mathrm{0}},T}\left(E_{n}\right)\left(1-F_{\mu_{\mathrm{0}},T}\left(\xi_{k}\right)\right)$
is explicitly rewritten as
\begin{align}
I_{\mathrm{-,inela}} & =2\pi e\int dE_{n}\rho_{\mathrm{s}}\left(E_{n}\right)\rho_{\mathrm{t}}\left(\xi_{k}\right)F_{\mu_{\mathrm{0}},T}\left(E_{n}\right)\nonumber \\
 & \times\left(1-F_{\mu_{\mathrm{0}},T}\left(\xi_{k}\right)\right)\left|\mathscr{N}_{\mathrm{s,t}}|_{V_{\mathrm{b}},E_{n}\rightarrow\xi_{k}}\right|^{2}|_{\xi_{k}=E_{n}-eV_{\mathrm{b}}-E_{eg}},
\end{align}
where $\rho_{\mathrm{t}}\left(E\right)$ ($\rho_{\mathrm{s}}\left(E\right)$)
are the density of state of tip (substrate) at the energy $E$. $F_{\mu_{\mathrm{0},}T}\left(E\right)$
is the Fermi-Dirac distribution of electrons in tip or substrate state
at energy $E$, chemical potential $\mu_{\mathrm{0}}$, and temperature
$T$.\textcolor{red}{{} }$\left|\mathscr{N}_{\mathrm{s,t}}|_{V_{\mathrm{b}},E_{n}\rightarrow\xi_{k}}\right|^{2}|_{\xi_{k}=E_{n}-eV_{\mathrm{b}}-E_{eg}}$
rules out all the tunneling processes whose energy do not conserve.
\textcolor{black}{Without loss of generality, we consider here the
tip and substrate are of the same metal (Ag).}

In STML experiment, the temperature of the ultrahigh-vacuum chamber
is low enough, typically lower than 10K \citep{GongChen_PRL_2019_triplet_up_conversion,YangZhang_Nature_2016_dipole_dipole,Imada_nature_2016_energy_transfer,KensukeKimura_nature_triplet_2019,Doppagne_science_2018,YangLuo_Single_photon_superradiance_PRL_2019,HiroshiImada_fano_PRL_2017,YaoZhang_NC_fano_lineshape_2017,Doppagne_PRL_2017,Kroger_nanolett_fano_2018},
that the Fermi-Dirac distribution function is approximately a Heaviside
function, i.e., $F_{\mu_{0},T}\left(E\right)=1$ for $E<\mu_{0}$
and $F_{\mu_{0},T}\left(E\right)=0$ for $E>\mu_{0}$. The inelastic
tunneling current becomes

\begin{align}
I_{\mathrm{-,inela}} & \simeq2\pi e\int_{\mu_{0}+eV_{\mathrm{b}}+E_{eg}}^{\mu_{0}}dE_{n}\rho_{\mathrm{s}}\left(E_{n}\right)\rho_{\mathrm{t}}\left(\xi_{k}\right)\nonumber \\
 & \times\left|\mathscr{N}_{\mathrm{s,t}}|_{V_{\mathrm{b}},E_{n}\rightarrow\xi_{k}}\right|^{2}|_{\xi_{k}=E_{n}-eV_{\mathrm{b}}-E_{eg}}\label{eq:I1_inela}
\end{align}
Eq. (\ref{eq:I1_inela}) suggests that the current for inelastic tunneling
is nonzero only at the condition $eV_{\mathrm{b}}<-E_{eg}$ for the
negative bias case.

For the positive bias $V_{\mathrm{b}}>0$, the current for the inelastic
tunneling is obtained with the similar method as

\begin{align}
I_{\mathrm{+,inela}} & \simeq2\pi e\int_{\mu_{0}}^{\mu_{0}+eV_{b}-E_{eg}}dE_{n}\rho_{\mathrm{s}}\left(E_{n}\right)\rho_{\mathrm{t}}\left(\xi_{k}\right)\nonumber \\
 & \times\left|\mathscr{N}_{\mathrm{t,s}}|_{V_{\mathrm{b}},\xi_{k}\rightarrow E_{n}}\right|^{2}|_{\xi_{k}=E_{n}-eV_{\mathrm{b}}+E_{eg}}.\label{eq:I2_inela}
\end{align}
Similar to the negative bias case, the condition for a nonzero inelastic
current is $eV_{\mathrm{b}}>E_{eg}$. The equal bias voltage for nonzero
inelastic current at negative and positive bias is an important feature
different from the carrier-injection mechanism where the electron
injection requires different voltage for the negative and positive
bias \citep{YangZhang_Nature_2016_dipole_dipole,MichaelChong_thesis_2016}.
With Eqs. (\ref{eq:I1_inela} and \ref{eq:I2_inela}), we obtain the
inelastic tunneling current as
\begin{equation}
I_{\mathrm{inela}}=\begin{cases}
I_{\mathrm{-,inela}}, & V_{\mathrm{b}}<-\frac{E_{eg}}{e}\\
0, & -\frac{E_{eg}}{e}\leq V_{\mathrm{b}}\leq\frac{E_{eg}}{e}\\
I_{\mathrm{+,inela}}, & V_{\mathrm{b}}>\frac{E_{eg}}{e}.
\end{cases}.\label{eq:I_inela}
\end{equation}

Photon counting of molecular fluorescence is a quantity relevant for
probing the properties of the single molecule. Once excited, the molecule
will decay to its lower state spontaneously with rate $\gamma$. The
photon counting rate is proportional to the inelastic current
\begin{align}
\Gamma & =I_{\mathrm{inela}}/e.\label{eq:molecular_emission_intensity}
\end{align}
The detailed derivation can be found in the supplementary materials.
In Fig. \ref{fig:Photon-intensity}, we plot the photon counting rate
as the function of the bias voltage between the tip and the substrate.
The blue solid and black dashed lines show the relative emission intensity
for tip's center of curvature $R=0.5\mathrm{nm}$ and $1\mathrm{nm}$,
respectively. The Fermi energy of silver is $\mu_{0}=-4.64$eV, and
the density of state of silver can be found in \citep{handbook}.
Without loss of generality, we choose the tip right above the molecule
($a_{x}=0$) and the molecular dipole along the $z$ direction ($\mu_{z}\neq0$
while $\mu_{x}=\mu_{y}=0$). The distance between tip and molecule
is $d=0.4$nm. As predicted in Eq. (\ref{eq:I_inela}), the bias voltages
for nonzero inelastic current at negative and positive bias are the
same, i.e., $\left|eV_{\mathrm{b}}\right|>E_{eg}=2$eV. Insets in
Fig. \ref{fig:Photon-intensity} describe the mechanism of the inelastic
electron scattering.

\begin{figure}
\begin{centering}
\includegraphics[scale=0.38]{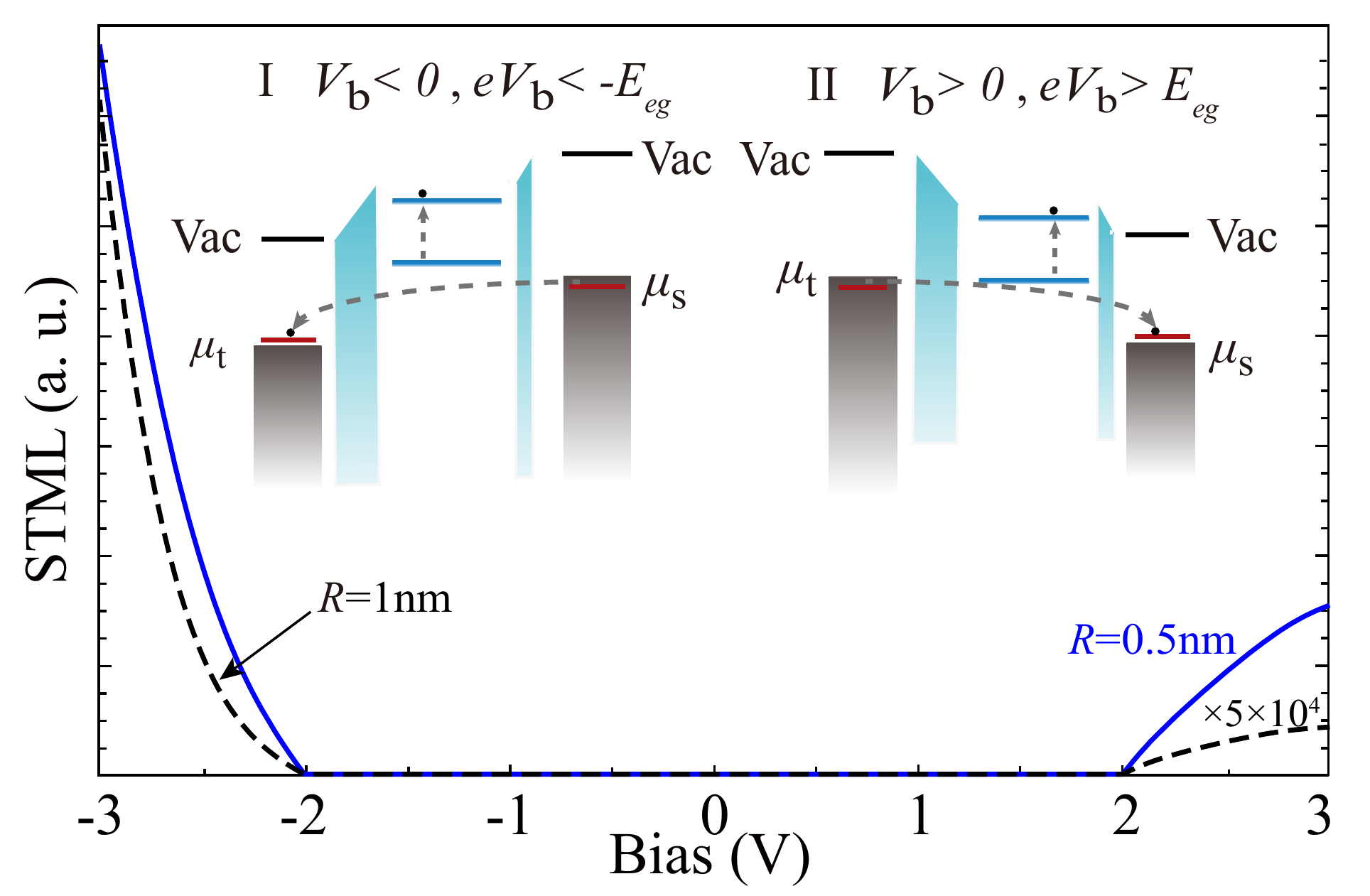}
\par\end{centering}
\caption{\label{fig:Photon-intensity}Asymmetric photon intensity in inelastic
electron scattering mechanism. The blue solid (black dashed) line
represents the emission intensity for tip's center of curvature $R=0.5\left(1\right)$nm.
Two insets show the inelastic electron scattering mechanism for a
two-level molecule at the negative and positive bias. The molecular
optical gap is $E_{eg}=2$eV, and the Fermi energy of silver is $\mu_{0}=-4.64$eV.
The distance between tip and molecule is $d=0.4$nm. Here, we choose
the case where the tip is deposited right above the molecule and the
molecular transition dipole is along the $z$ axis.}
\end{figure}

Another important feature is the asymmetry of the larger photon counting
at negative bias than that at positive bias, as illustrated in Fig.
\ref{fig:Photon-intensity}. This intensity asymmetry stems from the
eigenfunction asymmetry of tip and substrate. The tip's wave function
$\phi_{k}\left(\vec{\mathbf{r}}\right)$ decays spherically with factor
$\kappa_{k}$, and substrate's wave function $\varphi_{n}\left(\vec{\mathbf{r}}\right)$
decays along the $+z$ direction with factor $\kappa_{n}$. The relation
between the elements of the transition matrix at positive bias $V_{\mathrm{b}}$
and that at negative bias $-V_{\mathrm{b}}$ reads

\begin{align}
\mathscr{N}_{\mathrm{t,s}}|_{V_{\mathrm{b}},\xi_{k}\rightarrow E_{n}} & \simeq e^{-\left(\kappa_{k}-\kappa_{n}\right)R}\mathcal{N}_{\mathrm{s,t}}|_{-V_{\mathrm{b}},\xi_{k}\rightarrow E_{n}}.\label{eq:transition_element_relation}
\end{align}
The ratio between the transition matrix element at positive bias $V_{\mathrm{b}}$
and that at negative bias $-V_{\mathrm{b}}$ is $e^{-\left(\kappa_{k}-\kappa_{n}\right)R}$
. Inserting Eq. (\ref{eq:transition_element_relation}) into Eq. (\ref{eq:I1_inela}),
we obtain the ratio of the emission intensity as (see Supplementary
Material for details)
\begin{equation}
\mathscr{R}=\frac{I_{+\mathrm{,inela}}|_{V_{\mathrm{b}}}}{I_{\mathrm{-,inela}}|_{-V_{\mathrm{b}}}}\simeq e^{-R\sqrt{-2m_{e}\mu_{0}}\frac{eV_{\mathrm{b}}-E_{eg}}{-\mu_{0}}}<1.\label{eq:ratio}
\end{equation}
The current equation shows the characteristic asymmetry with larger
current at negative bias induced by inelastic electron tunneling.
Such asymmetry for inelastic scattering is caused by geometry shape
of the tip and the substrate, and persists with different materials.

In Fig. 3, we show the dependence of the asymmetrical ratio $\mathscr{R}$
as a function of the bias voltage with both the analytical formula
(red dashed line) in Eq. (\ref{eq:ratio}) and the numerical result
(blue solid line) calculated with the exact tunneling rate from Eqs.
(\ref{eq:I1_inela}-\ref{eq:I2_inela}). The analytical formula shows
an agreement on the trend that the asymmetry of the photon counting
increases with increasing bias voltage. The exponential decay of the
ratio $\mathscr{R}$ as function of bias voltage is predicted in Eq.
(\ref{eq:ratio}) and shall be tested with the experimental data.

\begin{figure}
\begin{centering}
\includegraphics[scale=0.35]{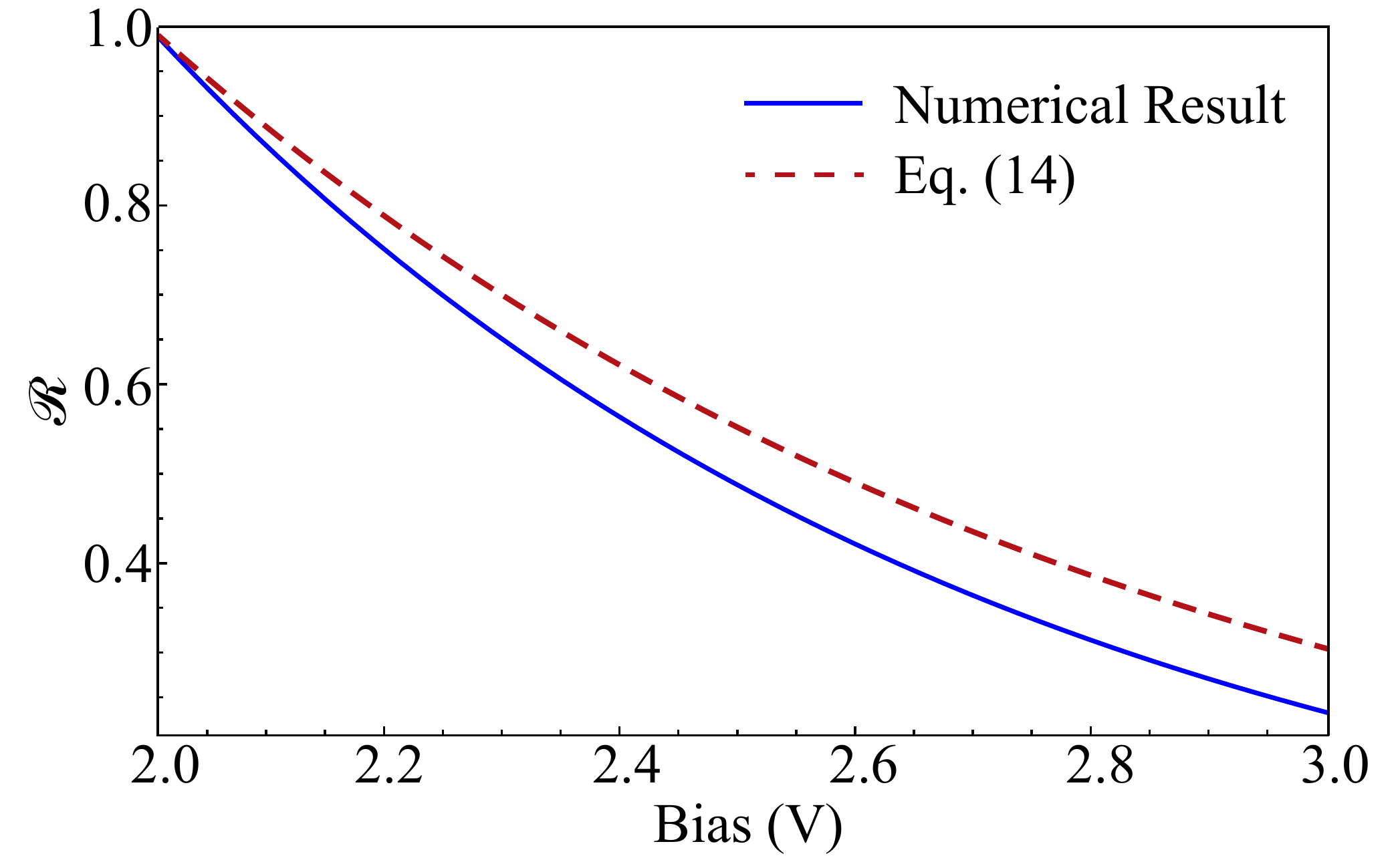}
\par\end{centering}
\caption{The ratio $\mathscr{R}$ between the photon counting at the positive
and negative voltages. The solid line shows the ratio calculated with
the exact tunneling rate from Eqs. (\ref{eq:I1_inela} and \ref{eq:I2_inela}),
and the dashed line shows the analytical formula for the ratio in
Eq. (\ref{eq:ratio}).}

\end{figure}

With the theoretical predictions above, we revisit the important features
observed in recent experiments \citep{YangZhang_Nature_2016_dipole_dipole,Doppagne_science_2018,Kroger_nanolett_fano_2018}.
In the single-hydrocarbon fluorescence induced by STM \citep{Kroger_nanolett_fano_2018},
the phenomenon that the emission intensity at positive bias was lower
than that at negative bias is in line with our prediction. Though
such the asymmetric intensity feature (the intensity at positive bias
was much lower than that at negative bias) of a single ZnPc molecule
was attributed to the carrier-injection mechanism \citep{YangZhang_Nature_2016_dipole_dipole},
we emphasis that the inelastic electron scattering mechanism may also
play an important role in this feature. By changing the tip and substrate
material from Ag to Au, Doppagne \textit{el al.} \citep{Doppagne_science_2018}
observed a phenomenon which was opposite to the feature in \citep{YangZhang_Nature_2016_dipole_dipole}.
The emission of a single neutral ZnPc molecule at positive bias was
30 times more intense than that at negative bias. Our theory definitely
excludes the inelastic electron scattering mechanism as the origin
of such asymmetric luminescence in \citep{Doppagne_science_2018}.

In conclusion, we have derived the microscopic origin of the molecular
excitation via the inelastic electron scattering mechanism in single-molecule
STML. By the model, we obtain the emission intensity in the inelastic
electron scattering mechanism. We find that inelastic electron scattering
mechanism requires a symmetric bias voltage for nonzero inelastic
current which equals the optical gap of this two-level molecule exactly.
It implies that the energy window between the Fermi levels of two
electrodes should at least equal the optical gap of the molecule \citep{MichaelChong_thesis_2016}.
Importantly, we reveal an asymmetric emission intensity at negative
and positive bias which is due to the asymmetric forms of wave functions
at two electrodes and show that the ratio of such asymmetry decays
with tip's radius of curvature and bias voltage. Our model offers
us a theoretical insight into the molecular excitation in the inelastic
electron scattering process which has never been explored before.

Before closing, it is worthy to mention that the inelastic scattering
mechanism is one of the three mechanisms proposed now and the photon
counting obtained here is one part of the total emission intensity.
Further research is needed for elucidating the competition of these
three mechanisms and finding the dominant one under certain conditions.
\begin{acknowledgments}
H. D. thanks Yang Zhang for the helpful discussion. This work is supported
 by the NSFC (Grants No. 11534002 and No. 11875049),
the NSAF (Grant No. U1730449 and No. U1530401), and the National Basic
Research Program of China (Grants No. 2016YFA0301201 and No. 2014CB921403).
H.D. also thanks The Recruitment Program of Global Youth Experts of
China.
\end{acknowledgments}

\bibliographystyle{apsrev4-1}

\begin{thebibliography}{27}%
\makeatletter
\providecommand \@ifxundefined [1]{%
 \@ifx{#1\undefined}
}%
\providecommand \@ifnum [1]{%
 \ifnum #1\expandafter \@firstoftwo
 \else \expandafter \@secondoftwo
 \fi
}%
\providecommand \@ifx [1]{%
 \ifx #1\expandafter \@firstoftwo
 \else \expandafter \@secondoftwo
 \fi
}%
\providecommand \natexlab [1]{#1}%
\providecommand \enquote  [1]{``#1''}%
\providecommand \bibnamefont  [1]{#1}%
\providecommand \bibfnamefont [1]{#1}%
\providecommand \citenamefont [1]{#1}%
\providecommand \href@noop [0]{\@secondoftwo}%
\providecommand \href [0]{\begingroup \@sanitize@url \@href}%
\providecommand \@href[1]{\@@startlink{#1}\@@href}%
\providecommand \@@href[1]{\endgroup#1\@@endlink}%
\providecommand \@sanitize@url [0]{\catcode `\\12\catcode `\$12\catcode
  `\&12\catcode `\#12\catcode `\^12\catcode `\_12\catcode `\%12\relax}%
\providecommand \@@startlink[1]{}%
\providecommand \@@endlink[0]{}%
\providecommand \url  [0]{\begingroup\@sanitize@url \@url }%
\providecommand \@url [1]{\endgroup\@href {#1}{\urlprefix }}%
\providecommand \urlprefix  [0]{URL }%
\providecommand \Eprint [0]{\href }%
\providecommand \doibase [0]{http://dx.doi.org/}%
\providecommand \selectlanguage [0]{\@gobble}%
\providecommand \bibinfo  [0]{\@secondoftwo}%
\providecommand \bibfield  [0]{\@secondoftwo}%
\providecommand \translation [1]{[#1]}%
\providecommand \BibitemOpen [0]{}%
\providecommand \bibitemStop [0]{}%
\providecommand \bibitemNoStop [0]{.\EOS\space}%
\providecommand \EOS [0]{\spacefactor3000\relax}%
\providecommand \BibitemShut  [1]{\csname bibitem#1\endcsname}%
\let\auto@bib@innerbib\@empty
\bibitem [{\citenamefont {Aradhya}\ and\ \citenamefont
  {Venkataraman}(2013)}]{Aradhya_2013}%
  \BibitemOpen
  \bibfield  {author} {\bibinfo {author} {\bibfnamefont {S.~V.}\ \bibnamefont
  {Aradhya}}\ and\ \bibinfo {author} {\bibfnamefont {L.}~\bibnamefont
  {Venkataraman}},\ }\href {\doibase 10.1038/nnano.2013.91} {\bibfield
  {journal} {\bibinfo  {journal} {Nat. Nanotechnol.}\ }\textbf {\bibinfo
  {volume} {8}},\ \bibinfo {pages} {399} (\bibinfo {year} {2013})}\BibitemShut
  {NoStop}%
\bibitem [{\citenamefont {Xin}\ \emph {et~al.}(2019)\citenamefont {Xin},
  \citenamefont {Guan}, \citenamefont {Zhou}, \citenamefont {Chen},
  \citenamefont {Gu}, \citenamefont {Li}, \citenamefont {Ratner}, \citenamefont
  {Nitzan}, \citenamefont {Stoddart},\ and\ \citenamefont {Guo}}]{XinNa_2019}%
  \BibitemOpen
  \bibfield  {author} {\bibinfo {author} {\bibfnamefont {N.}~\bibnamefont
  {Xin}}, \bibinfo {author} {\bibfnamefont {J.}~\bibnamefont {Guan}}, \bibinfo
  {author} {\bibfnamefont {C.}~\bibnamefont {Zhou}}, \bibinfo {author}
  {\bibfnamefont {X.}~\bibnamefont {Chen}}, \bibinfo {author} {\bibfnamefont
  {C.}~\bibnamefont {Gu}}, \bibinfo {author} {\bibfnamefont {Y.}~\bibnamefont
  {Li}}, \bibinfo {author} {\bibfnamefont {M.~A.}\ \bibnamefont {Ratner}},
  \bibinfo {author} {\bibfnamefont {A.}~\bibnamefont {Nitzan}}, \bibinfo
  {author} {\bibfnamefont {J.~F.}\ \bibnamefont {Stoddart}}, \ and\ \bibinfo
  {author} {\bibfnamefont {X.}~\bibnamefont {Guo}},\ }\href {\doibase
  10.1038/s42254-019-0022-x} {\bibfield  {journal} {\bibinfo  {journal} {Nat.
  Rev. Phys.}\ }\textbf {\bibinfo {volume} {1}},\ \bibinfo {pages} {211}
  (\bibinfo {year} {2019})}\BibitemShut {NoStop}%
\bibitem [{\citenamefont {Sun}\ \emph {et~al.}(2014)\citenamefont {Sun},
  \citenamefont {Diaz-Fernandez}, \citenamefont {Gschneidtner}, \citenamefont
  {Westerlund}, \citenamefont {Lara-Avila},\ and\ \citenamefont
  {Moth-Poulsen}}]{sunlanlan_single_molecule_2014}%
  \BibitemOpen
  \bibfield  {author} {\bibinfo {author} {\bibfnamefont {L.~L.}\ \bibnamefont
  {Sun}}, \bibinfo {author} {\bibfnamefont {Y.~A.}\ \bibnamefont
  {Diaz-Fernandez}}, \bibinfo {author} {\bibfnamefont {T.~A.}\ \bibnamefont
  {Gschneidtner}}, \bibinfo {author} {\bibfnamefont {F.}~\bibnamefont
  {Westerlund}}, \bibinfo {author} {\bibfnamefont {S.}~\bibnamefont
  {Lara-Avila}}, \ and\ \bibinfo {author} {\bibfnamefont {K.}~\bibnamefont
  {Moth-Poulsen}},\ }\href {\doibase 10.1039/c4cs00143e} {\bibfield  {journal}
  {\bibinfo  {journal} {Chem. Soc. Rev.}\ }\textbf {\bibinfo {volume} {43}},\
  \bibinfo {pages} {7378} (\bibinfo {year} {2014})}\BibitemShut {NoStop}%
\bibitem [{\citenamefont {Flaxer}\ \emph {et~al.}(1993)\citenamefont {Flaxer},
  \citenamefont {Sneh},\ and\ \citenamefont {Cheshnovsky}}]{Eli_science_1993}%
  \BibitemOpen
  \bibfield  {author} {\bibinfo {author} {\bibfnamefont {E.}~\bibnamefont
  {Flaxer}}, \bibinfo {author} {\bibfnamefont {O.}~\bibnamefont {Sneh}}, \ and\
  \bibinfo {author} {\bibfnamefont {O.}~\bibnamefont {Cheshnovsky}},\ }\href
  {\doibase 10.1126/science.262.5142.2012} {\bibfield  {journal} {\bibinfo
  {journal} {Science}\ }\textbf {\bibinfo {volume} {262}},\ \bibinfo {pages}
  {2012} (\bibinfo {year} {1993})}\BibitemShut {NoStop}%
\bibitem [{\citenamefont {Berndt}\ \emph {et~al.}(1993)\citenamefont {Berndt},
  \citenamefont {Gaisch}, \citenamefont {Gimzewski}, \citenamefont {Reihi},
  \citenamefont {Schlittler}, \citenamefont {Schneider},\ and\ \citenamefont
  {Tschudy}}]{Berndt_science_1993}%
  \BibitemOpen
  \bibfield  {author} {\bibinfo {author} {\bibfnamefont {R.}~\bibnamefont
  {Berndt}}, \bibinfo {author} {\bibfnamefont {R.}~\bibnamefont {Gaisch}},
  \bibinfo {author} {\bibfnamefont {J.~K.}\ \bibnamefont {Gimzewski}}, \bibinfo
  {author} {\bibfnamefont {B.}~\bibnamefont {Reihi}}, \bibinfo {author}
  {\bibfnamefont {R.~R.}\ \bibnamefont {Schlittler}}, \bibinfo {author}
  {\bibfnamefont {W.~D.}\ \bibnamefont {Schneider}}, \ and\ \bibinfo {author}
  {\bibfnamefont {M.}~\bibnamefont {Tschudy}},\ }\href {\doibase
  10.1126/science.262.5138.1425} {\bibfield  {journal} {\bibinfo  {journal}
  {Science}\ }\textbf {\bibinfo {volume} {262}},\ \bibinfo {pages} {1425}
  (\bibinfo {year} {1993})}\BibitemShut {NoStop}%
\bibitem [{\citenamefont {Zhang}\ \emph {et~al.}(2016)\citenamefont {Zhang},
  \citenamefont {Luo}, \citenamefont {Zhang}, \citenamefont {Yu}, \citenamefont
  {Kuang}, \citenamefont {Zhang}, \citenamefont {Meng}, \citenamefont {Luo},
  \citenamefont {Yang}, \citenamefont {Dong},\ and\ \citenamefont
  {Hou}}]{YangZhang_Nature_2016_dipole_dipole}%
  \BibitemOpen
  \bibfield  {author} {\bibinfo {author} {\bibfnamefont {Y.}~\bibnamefont
  {Zhang}}, \bibinfo {author} {\bibfnamefont {Y.}~\bibnamefont {Luo}}, \bibinfo
  {author} {\bibfnamefont {Y.}~\bibnamefont {Zhang}}, \bibinfo {author}
  {\bibfnamefont {Y.~J.}\ \bibnamefont {Yu}}, \bibinfo {author} {\bibfnamefont
  {Y.~M.}\ \bibnamefont {Kuang}}, \bibinfo {author} {\bibfnamefont
  {L.}~\bibnamefont {Zhang}}, \bibinfo {author} {\bibfnamefont {Q.~S.}\
  \bibnamefont {Meng}}, \bibinfo {author} {\bibfnamefont {Y.}~\bibnamefont
  {Luo}}, \bibinfo {author} {\bibfnamefont {J.~L.}\ \bibnamefont {Yang}},
  \bibinfo {author} {\bibfnamefont {Z.~C.}\ \bibnamefont {Dong}}, \ and\
  \bibinfo {author} {\bibfnamefont {J.~G.}\ \bibnamefont {Hou}},\ }\href
  {\doibase 10.1038/nature17428} {\bibfield  {journal} {\bibinfo  {journal}
  {Nature (London)}\ }\textbf {\bibinfo {volume} {531}},\ \bibinfo {pages}
  {623} (\bibinfo {year} {2016})}\BibitemShut {NoStop}%
\bibitem [{\citenamefont {Doppagne}\ \emph {et~al.}(2017)\citenamefont
  {Doppagne}, \citenamefont {Chong}, \citenamefont {Lorchat}, \citenamefont
  {Berciaud}, \citenamefont {Romeo}, \citenamefont {Bulou}, \citenamefont
  {Boeglin}, \citenamefont {Scheurer},\ and\ \citenamefont
  {Schull}}]{Doppagne_PRL_2017}%
  \BibitemOpen
  \bibfield  {author} {\bibinfo {author} {\bibfnamefont {B.}~\bibnamefont
  {Doppagne}}, \bibinfo {author} {\bibfnamefont {M.~C.}\ \bibnamefont {Chong}},
  \bibinfo {author} {\bibfnamefont {E.}~\bibnamefont {Lorchat}}, \bibinfo
  {author} {\bibfnamefont {S.}~\bibnamefont {Berciaud}}, \bibinfo {author}
  {\bibfnamefont {M.}~\bibnamefont {Romeo}}, \bibinfo {author} {\bibfnamefont
  {H.}~\bibnamefont {Bulou}}, \bibinfo {author} {\bibfnamefont
  {A.}~\bibnamefont {Boeglin}}, \bibinfo {author} {\bibfnamefont
  {F.}~\bibnamefont {Scheurer}}, \ and\ \bibinfo {author} {\bibfnamefont
  {G.}~\bibnamefont {Schull}},\ }\href {\doibase
  10.1103/PhysRevLett.118.127401} {\bibfield  {journal} {\bibinfo  {journal}
  {Phys. Rev. Lett.}\ }\textbf {\bibinfo {volume} {118}},\ \bibinfo {pages}
  {127401} (\bibinfo {year} {2017})}\BibitemShut {NoStop}%
\bibitem [{\citenamefont {Luo}\ \emph {et~al.}(2019)\citenamefont {Luo},
  \citenamefont {Chen}, \citenamefont {Zhang}, \citenamefont {Zhang},
  \citenamefont {Yu}, \citenamefont {Kong}, \citenamefont {Tian}, \citenamefont
  {Zhang}, \citenamefont {Shan}, \citenamefont {Luo}, \citenamefont {Yang},
  \citenamefont {Sandoghdar}, \citenamefont {Dong},\ and\ \citenamefont
  {Hou}}]{YangLuo_Single_photon_superradiance_PRL_2019}%
  \BibitemOpen
  \bibfield  {author} {\bibinfo {author} {\bibfnamefont {Y.}~\bibnamefont
  {Luo}}, \bibinfo {author} {\bibfnamefont {G.}~\bibnamefont {Chen}}, \bibinfo
  {author} {\bibfnamefont {Y.}~\bibnamefont {Zhang}}, \bibinfo {author}
  {\bibfnamefont {L.}~\bibnamefont {Zhang}}, \bibinfo {author} {\bibfnamefont
  {Y.~J.}\ \bibnamefont {Yu}}, \bibinfo {author} {\bibfnamefont {F.~F.}\
  \bibnamefont {Kong}}, \bibinfo {author} {\bibfnamefont {X.~J.}\ \bibnamefont
  {Tian}}, \bibinfo {author} {\bibfnamefont {Y.}~\bibnamefont {Zhang}},
  \bibinfo {author} {\bibfnamefont {C.~X.}\ \bibnamefont {Shan}}, \bibinfo
  {author} {\bibfnamefont {Y.}~\bibnamefont {Luo}}, \bibinfo {author}
  {\bibfnamefont {J.~L.}\ \bibnamefont {Yang}}, \bibinfo {author}
  {\bibfnamefont {V.}~\bibnamefont {Sandoghdar}}, \bibinfo {author}
  {\bibfnamefont {Z.~C.}\ \bibnamefont {Dong}}, \ and\ \bibinfo {author}
  {\bibfnamefont {J.~G.}\ \bibnamefont {Hou}},\ }\href {\doibase
  10.1103/PhysRevLett.122.233901} {\bibfield  {journal} {\bibinfo  {journal}
  {Phys. Rev. Lett.}\ }\textbf {\bibinfo {volume} {122}},\ \bibinfo {pages}
  {233901} (\bibinfo {year} {2019})}\BibitemShut {NoStop}%
\bibitem [{\citenamefont {Imada}\ \emph {et~al.}(2016)\citenamefont {Imada},
  \citenamefont {Miwa}, \citenamefont {Imai-Imada}, \citenamefont {Kawahara},
  \citenamefont {Kimura},\ and\ \citenamefont
  {Kim}}]{Imada_nature_2016_energy_transfer}%
  \BibitemOpen
  \bibfield  {author} {\bibinfo {author} {\bibfnamefont {H.}~\bibnamefont
  {Imada}}, \bibinfo {author} {\bibfnamefont {K.}~\bibnamefont {Miwa}},
  \bibinfo {author} {\bibfnamefont {M.}~\bibnamefont {Imai-Imada}}, \bibinfo
  {author} {\bibfnamefont {S.}~\bibnamefont {Kawahara}}, \bibinfo {author}
  {\bibfnamefont {K.}~\bibnamefont {Kimura}}, \ and\ \bibinfo {author}
  {\bibfnamefont {Y.}~\bibnamefont {Kim}},\ }\href {\doibase
  10.1038/nature19765} {\bibfield  {journal} {\bibinfo  {journal} {Nature
  (London)}\ }\textbf {\bibinfo {volume} {538}},\ \bibinfo {pages} {364}
  (\bibinfo {year} {2016})}\BibitemShut {NoStop}%
\bibitem [{\citenamefont {Imada}\ \emph {et~al.}(2017)\citenamefont {Imada},
  \citenamefont {Miwa}, \citenamefont {Imai-Imada}, \citenamefont {Kawahara},
  \citenamefont {Kimura},\ and\ \citenamefont
  {Kim}}]{HiroshiImada_fano_PRL_2017}%
  \BibitemOpen
  \bibfield  {author} {\bibinfo {author} {\bibfnamefont {H.}~\bibnamefont
  {Imada}}, \bibinfo {author} {\bibfnamefont {K.}~\bibnamefont {Miwa}},
  \bibinfo {author} {\bibfnamefont {M.}~\bibnamefont {Imai-Imada}}, \bibinfo
  {author} {\bibfnamefont {S.}~\bibnamefont {Kawahara}}, \bibinfo {author}
  {\bibfnamefont {K.}~\bibnamefont {Kimura}}, \ and\ \bibinfo {author}
  {\bibfnamefont {Y.}~\bibnamefont {Kim}},\ }\href {\doibase
  10.1103/PhysRevLett.119.013901} {\bibfield  {journal} {\bibinfo  {journal}
  {Phys. Rev. Lett.}\ }\textbf {\bibinfo {volume} {119}},\ \bibinfo {pages}
  {013901} (\bibinfo {year} {2017})}\BibitemShut {NoStop}%
\bibitem [{\citenamefont {Zhang}\ \emph {et~al.}(2017)\citenamefont {Zhang},
  \citenamefont {Meng}, \citenamefont {Zhang}, \citenamefont {Luo},
  \citenamefont {Yu}, \citenamefont {Yang}, \citenamefont {Zhang},
  \citenamefont {Esteban}, \citenamefont {Aizpurua}, \citenamefont {Luo},
  \citenamefont {Yang}, \citenamefont {Dong},\ and\ \citenamefont
  {Hou}}]{YaoZhang_NC_fano_lineshape_2017}%
  \BibitemOpen
  \bibfield  {author} {\bibinfo {author} {\bibfnamefont {Y.}~\bibnamefont
  {Zhang}}, \bibinfo {author} {\bibfnamefont {Q.~S.}\ \bibnamefont {Meng}},
  \bibinfo {author} {\bibfnamefont {L.}~\bibnamefont {Zhang}}, \bibinfo
  {author} {\bibfnamefont {Y.}~\bibnamefont {Luo}}, \bibinfo {author}
  {\bibfnamefont {Y.~J.}\ \bibnamefont {Yu}}, \bibinfo {author} {\bibfnamefont
  {B.}~\bibnamefont {Yang}}, \bibinfo {author} {\bibfnamefont {Y.}~\bibnamefont
  {Zhang}}, \bibinfo {author} {\bibfnamefont {R.}~\bibnamefont {Esteban}},
  \bibinfo {author} {\bibfnamefont {J.}~\bibnamefont {Aizpurua}}, \bibinfo
  {author} {\bibfnamefont {Y.}~\bibnamefont {Luo}}, \bibinfo {author}
  {\bibfnamefont {J.~L.}\ \bibnamefont {Yang}}, \bibinfo {author}
  {\bibfnamefont {Z.~C.}\ \bibnamefont {Dong}}, \ and\ \bibinfo {author}
  {\bibfnamefont {J.~G.}\ \bibnamefont {Hou}},\ }\href {\doibase
  10.1038/ncomms15225} {\bibfield  {journal} {\bibinfo  {journal} {Nat.
  Commun.}\ }\textbf {\bibinfo {volume} {8}},\ \bibinfo {pages} {15225}
  (\bibinfo {year} {2017})}\BibitemShut {NoStop}%
\bibitem [{\citenamefont {Kr\"oger}\ \emph {et~al.}(2018)\citenamefont
  {Kr\"oger}, \citenamefont {Doppagne}, \citenamefont {Scheurer},\ and\
  \citenamefont {Schull}}]{Kroger_nanolett_fano_2018}%
  \BibitemOpen
  \bibfield  {author} {\bibinfo {author} {\bibfnamefont {J.}~\bibnamefont
  {Kr\"oger}}, \bibinfo {author} {\bibfnamefont {B.}~\bibnamefont {Doppagne}},
  \bibinfo {author} {\bibfnamefont {F.}~\bibnamefont {Scheurer}}, \ and\
  \bibinfo {author} {\bibfnamefont {G.}~\bibnamefont {Schull}},\ }\href
  {\doibase 10.1021/acs.nanolett.8b00304} {\bibfield  {journal} {\bibinfo
  {journal} {Nano Lett.}\ }\textbf {\bibinfo {volume} {18}},\ \bibinfo {pages}
  {3407} (\bibinfo {year} {2018})}\BibitemShut {NoStop}%
\bibitem [{\citenamefont {Doppagne}\ \emph {et~al.}(2018)\citenamefont
  {Doppagne}, \citenamefont {Chong}, \citenamefont {Bulou}, \citenamefont
  {Boeglin}, \citenamefont {Scheurer},\ and\ \citenamefont
  {Schull}}]{Doppagne_science_2018}%
  \BibitemOpen
  \bibfield  {author} {\bibinfo {author} {\bibfnamefont {B.}~\bibnamefont
  {Doppagne}}, \bibinfo {author} {\bibfnamefont {M.~C.}\ \bibnamefont {Chong}},
  \bibinfo {author} {\bibfnamefont {H.}~\bibnamefont {Bulou}}, \bibinfo
  {author} {\bibfnamefont {A.}~\bibnamefont {Boeglin}}, \bibinfo {author}
  {\bibfnamefont {F.}~\bibnamefont {Scheurer}}, \ and\ \bibinfo {author}
  {\bibfnamefont {G.}~\bibnamefont {Schull}},\ }\href {\doibase
  10.1126/science.aat1603} {\bibfield  {journal} {\bibinfo  {journal}
  {Science}\ }\textbf {\bibinfo {volume} {361}},\ \bibinfo {pages} {251}
  (\bibinfo {year} {2018})}\BibitemShut {NoStop}%
\bibitem [{\citenamefont {Wu}\ \emph {et~al.}(2019)\citenamefont {Wu},
  \citenamefont {Wang}, \citenamefont {Zhang}, \citenamefont {Song},\ and\
  \citenamefont {Yam}}]{simulation_xyWu_2019}%
  \BibitemOpen
  \bibfield  {author} {\bibinfo {author} {\bibfnamefont {X.~Y.}\ \bibnamefont
  {Wu}}, \bibinfo {author} {\bibfnamefont {R.~L.}\ \bibnamefont {Wang}},
  \bibinfo {author} {\bibfnamefont {Y.}~\bibnamefont {Zhang}}, \bibinfo
  {author} {\bibfnamefont {B.~W.}\ \bibnamefont {Song}}, \ and\ \bibinfo
  {author} {\bibfnamefont {C.~Y.}\ \bibnamefont {Yam}},\ }\href {\doibase
  10.1021/acs.jpcc.9b02198} {\bibfield  {journal} {\bibinfo  {journal} {J.
  Phys. Chem. C}\ }\textbf {\bibinfo {volume} {123}},\ \bibinfo {pages} {15761}
  (\bibinfo {year} {2019})}\BibitemShut {NoStop}%
\bibitem [{\citenamefont {Miwa}\ \emph
  {et~al.}(2019{\natexlab{a}})\citenamefont {Miwa}, \citenamefont {Imada},
  \citenamefont {Imai-Imada}, \citenamefont {Kimura}, \citenamefont
  {Galperin},\ and\ \citenamefont {Kim}}]{simulation_Miwa_2019}%
  \BibitemOpen
  \bibfield  {author} {\bibinfo {author} {\bibfnamefont {K.}~\bibnamefont
  {Miwa}}, \bibinfo {author} {\bibfnamefont {H.}~\bibnamefont {Imada}},
  \bibinfo {author} {\bibfnamefont {M.}~\bibnamefont {Imai-Imada}}, \bibinfo
  {author} {\bibfnamefont {K.}~\bibnamefont {Kimura}}, \bibinfo {author}
  {\bibfnamefont {M.}~\bibnamefont {Galperin}}, \ and\ \bibinfo {author}
  {\bibfnamefont {Y.}~\bibnamefont {Kim}},\ }\href {\doibase
  10.1021/acs.nanolett.8b04484} {\bibfield  {journal} {\bibinfo  {journal}
  {Nano Lett.}\ }\textbf {\bibinfo {volume} {19}},\ \bibinfo {pages} {2803}
  (\bibinfo {year} {2019}{\natexlab{a}})}\BibitemShut {NoStop}%
\bibitem [{\citenamefont {Bardeen}(1961)}]{bardeen_1961}%
  \BibitemOpen
  \bibfield  {author} {\bibinfo {author} {\bibfnamefont {J.}~\bibnamefont
  {Bardeen}},\ }\href {\doibase 10.1103/PhysRevLett.6.57} {\bibfield  {journal}
  {\bibinfo  {journal} {Phys. Rev. Lett.}\ }\textbf {\bibinfo {volume} {6}},\
  \bibinfo {pages} {57} (\bibinfo {year} {1961})}\BibitemShut {NoStop}%
\bibitem [{\citenamefont {Gottlieb}\ and\ \citenamefont
  {Wesoloski}(2006)}]{TUTORIAL_2006}%
  \BibitemOpen
  \bibfield  {author} {\bibinfo {author} {\bibfnamefont {A.~D.}\ \bibnamefont
  {Gottlieb}}\ and\ \bibinfo {author} {\bibfnamefont {L.}~\bibnamefont
  {Wesoloski}},\ }\href {\doibase 10.1088/0957-4484/17/8/R01} {\bibfield
  {journal} {\bibinfo  {journal} {Nanotechnology}\ }\textbf {\bibinfo {volume}
  {17}},\ \bibinfo {pages} {R57} (\bibinfo {year} {2006})}\BibitemShut
  {NoStop}%
\bibitem [{\citenamefont {Nian}\ \emph {et~al.}(2018)\citenamefont {Nian},
  \citenamefont {Wang},\ and\ \citenamefont {L\"u}}]{Nian_simulation_2018}%
  \BibitemOpen
  \bibfield  {author} {\bibinfo {author} {\bibfnamefont {L.~L.}\ \bibnamefont
  {Nian}}, \bibinfo {author} {\bibfnamefont {Y.}~\bibnamefont {Wang}}, \ and\
  \bibinfo {author} {\bibfnamefont {J.~T.}\ \bibnamefont {L\"u}},\ }\href
  {\doibase 10.1021/acs.nanolett.8b02706} {\bibfield  {journal} {\bibinfo
  {journal} {Nano Lett.}\ }\textbf {\bibinfo {volume} {18}},\ \bibinfo {pages}
  {6826} (\bibinfo {year} {2018})}\BibitemShut {NoStop}%
\bibitem [{\citenamefont {Nian}\ and\ \citenamefont
  {L\"u}(2019)}]{Nian_simu_2019}%
  \BibitemOpen
  \bibfield  {author} {\bibinfo {author} {\bibfnamefont {L.~L.}\ \bibnamefont
  {Nian}}\ and\ \bibinfo {author} {\bibfnamefont {J.~T.}\ \bibnamefont
  {L\"u}},\ }\href {\doibase 10.1021/acs.jpcc.9b00132} {\bibfield  {journal}
  {\bibinfo  {journal} {J. Phys. Chem. C}\ }\textbf {\bibinfo {volume} {123}},\
  \bibinfo {pages} {18508} (\bibinfo {year} {2019})}\BibitemShut {NoStop}%
\bibitem [{\citenamefont {Minkin}\ \emph {et~al.}(1970)\citenamefont {Minkin},
  \citenamefont {Osipov},\ and\ \citenamefont
  {Zhdanov}}]{Minkin_Dipole_Moment_1970}%
  \BibitemOpen
  \bibfield  {author} {\bibinfo {author} {\bibfnamefont {V.~I.}\ \bibnamefont
  {Minkin}}, \bibinfo {author} {\bibfnamefont {O.~A.}\ \bibnamefont {Osipov}},
  \ and\ \bibinfo {author} {\bibfnamefont {Y.~A.}\ \bibnamefont {Zhdanov}},\
  }\href {\doibase 10.1007/978-1-4684-1770-8} {\emph {\bibinfo {title} {DIPOLE
  MOMENTS IN ORGANIC CHEMISTRY}}},\ \bibinfo {edition} {1st}\ ed.,\ edited by\
  \bibinfo {editor} {\bibfnamefont {W.~E.}\ \bibnamefont {Vaughan}}\ (\bibinfo
  {publisher} {Plenum Press, New York},\ \bibinfo {year} {1970})\ \bibinfo
  {note} {pp. 79}\BibitemShut {NoStop}%
\bibitem [{\citenamefont {Chen}(1990)}]{chen_1990}%
  \BibitemOpen
  \bibfield  {author} {\bibinfo {author} {\bibfnamefont {C.~J.}\ \bibnamefont
  {Chen}},\ }\href {\doibase 10.1103/PhysRevB.42.8841} {\bibfield  {journal}
  {\bibinfo  {journal} {Phys. Rev. B}\ }\textbf {\bibinfo {volume} {42}},\
  \bibinfo {pages} {8841} (\bibinfo {year} {1990})}\BibitemShut {NoStop}%
\bibitem [{\citenamefont {Tersoff}\ and\ \citenamefont
  {Hamann}(1983)}]{Tersoff_Hamann_1983}%
  \BibitemOpen
  \bibfield  {author} {\bibinfo {author} {\bibfnamefont {J.}~\bibnamefont
  {Tersoff}}\ and\ \bibinfo {author} {\bibfnamefont {D.~R.}\ \bibnamefont
  {Hamann}},\ }\href {\doibase 10.1103/PhysRevLett.50.1998} {\bibfield
  {journal} {\bibinfo  {journal} {Phys. Rev. Lett.}\ }\textbf {\bibinfo
  {volume} {50}},\ \bibinfo {pages} {1998} (\bibinfo {year}
  {1983})}\BibitemShut {NoStop}%
\bibitem [{\citenamefont {Tersoff}\ and\ \citenamefont
  {Hamann}(1985)}]{Tersoff_Hamann_1985}%
  \BibitemOpen
  \bibfield  {author} {\bibinfo {author} {\bibfnamefont {J.}~\bibnamefont
  {Tersoff}}\ and\ \bibinfo {author} {\bibfnamefont {D.~R.}\ \bibnamefont
  {Hamann}},\ }\href {\doibase 10.1103/PhysRevB.31.805} {\bibfield  {journal}
  {\bibinfo  {journal} {Phys. Rev. B}\ }\textbf {\bibinfo {volume} {31}},\
  \bibinfo {pages} {805} (\bibinfo {year} {1985})}\BibitemShut {NoStop}%
\bibitem [{\citenamefont {Chen}\ \emph {et~al.}(2019)\citenamefont {Chen},
  \citenamefont {Luo}, \citenamefont {Gao}, \citenamefont {Jiang},
  \citenamefont {Yu}, \citenamefont {Zhang}, \citenamefont {Zhang},
  \citenamefont {Li}, \citenamefont {Zhang},\ and\ \citenamefont
  {Dong}}]{GongChen_PRL_2019_triplet_up_conversion}%
  \BibitemOpen
  \bibfield  {author} {\bibinfo {author} {\bibfnamefont {G.}~\bibnamefont
  {Chen}}, \bibinfo {author} {\bibfnamefont {Y.}~\bibnamefont {Luo}}, \bibinfo
  {author} {\bibfnamefont {H.~Y.}\ \bibnamefont {Gao}}, \bibinfo {author}
  {\bibfnamefont {J.}~\bibnamefont {Jiang}}, \bibinfo {author} {\bibfnamefont
  {Y.~J.}\ \bibnamefont {Yu}}, \bibinfo {author} {\bibfnamefont
  {L.}~\bibnamefont {Zhang}}, \bibinfo {author} {\bibfnamefont
  {Y.}~\bibnamefont {Zhang}}, \bibinfo {author} {\bibfnamefont {X.~G.}\
  \bibnamefont {Li}}, \bibinfo {author} {\bibfnamefont {Z.~Y.}\ \bibnamefont
  {Zhang}}, \ and\ \bibinfo {author} {\bibfnamefont {Z.~C.}\ \bibnamefont
  {Dong}},\ }\href {\doibase 10.1103/PhysRevLett.122.177401} {\bibfield
  {journal} {\bibinfo  {journal} {Phys. Rev. Lett.}\ }\textbf {\bibinfo
  {volume} {122}},\ \bibinfo {pages} {177401} (\bibinfo {year}
  {2019})}\BibitemShut {NoStop}%
\bibitem [{\citenamefont {Miwa}\ \emph
  {et~al.}(2019{\natexlab{b}})\citenamefont {Miwa}, \citenamefont {Imada},
  \citenamefont {Imai-Imada}, \citenamefont {Kawahara}, \citenamefont {Takeya},
  \citenamefont {Kawai}, \citenamefont {Galperin},\ and\ \citenamefont
  {Kim}}]{KensukeKimura_nature_triplet_2019}%
  \BibitemOpen
  \bibfield  {author} {\bibinfo {author} {\bibfnamefont {K.}~\bibnamefont
  {Miwa}}, \bibinfo {author} {\bibfnamefont {H.}~\bibnamefont {Imada}},
  \bibinfo {author} {\bibfnamefont {M.}~\bibnamefont {Imai-Imada}}, \bibinfo
  {author} {\bibfnamefont {S.}~\bibnamefont {Kawahara}}, \bibinfo {author}
  {\bibfnamefont {J.}~\bibnamefont {Takeya}}, \bibinfo {author} {\bibfnamefont
  {M.}~\bibnamefont {Kawai}}, \bibinfo {author} {\bibfnamefont
  {M.}~\bibnamefont {Galperin}}, \ and\ \bibinfo {author} {\bibfnamefont
  {Y.}~\bibnamefont {Kim}},\ }\href {\doibase 10.1038/s41586-019-1284-2}
  {\bibfield  {journal} {\bibinfo  {journal} {Nature (London)}\ }\textbf
  {\bibinfo {volume} {570}},\ \bibinfo {pages} {210} (\bibinfo {year}
  {2019}{\natexlab{b}})}\BibitemShut {NoStop}%
\bibitem [{\citenamefont {Chong}(2016)}]{MichaelChong_thesis_2016}%
  \BibitemOpen
  \bibfield  {author} {\bibinfo {author} {\bibfnamefont {M.~C.}\ \bibnamefont
  {Chong}},\ }\emph {\bibinfo {title} {Electrically driven fluorescence of
  single molecule junctions}},\ \href
  {https://tel.archives-ouvertes.fr/tel-01468663} {Ph.D. thesis},\ \bibinfo
  {school} {Universit\'{e} de Strasbourg, France} (\bibinfo {year}
  {2016})\BibitemShut {NoStop}%
\bibitem [{\citenamefont {Papaconstantopoulos}(2015)}]{handbook}%
  \BibitemOpen
  \bibfield  {author} {\bibinfo {author} {\bibfnamefont {D.~A.}\ \bibnamefont
  {Papaconstantopoulos}},\ }\href {\doibase 10.1007/978-1-4419-8264-3} {\emph
  {\bibinfo {title} {Handbook of the band structure of elemental solids: from Z
  = 1 to Z = 112}}},\ \bibinfo {edition} {2nd}\ ed.\ (\bibinfo  {publisher}
  {Springer, New York},\ \bibinfo {year} {2015})\ \bibinfo {note} {pp.
  243}\BibitemShut {NoStop}%
\end{thebibliography}

\end{document}


\begin{center}
Supplementary Material for
\par\end{center}
\title{Microscopic origin of molecule excitation via inelastic electron scattering
in scanning tunneling microscope}
\author{Guohui Dong}
\affiliation{Graduate School of Chinese Academy of Engineering Physics, Beijing
100084, China}
\author{Yining You}
\affiliation{Department of Modern Physics, University of Science and Technology
of China, Hefei 230026, China}
\author{Hui Dong}
\email{hdong@gscaep.ac.cn}

\affiliation{Graduate School of Chinese Academy of Engineering Physics, Beijing
100084, China}

\maketitle
This document is devoted to providing the detailed derivations and
the supporting discussions to the main content.

\section{Electronic wave functions on the tip and substrate}

\begin{figure}
\begin{centering}
\includegraphics{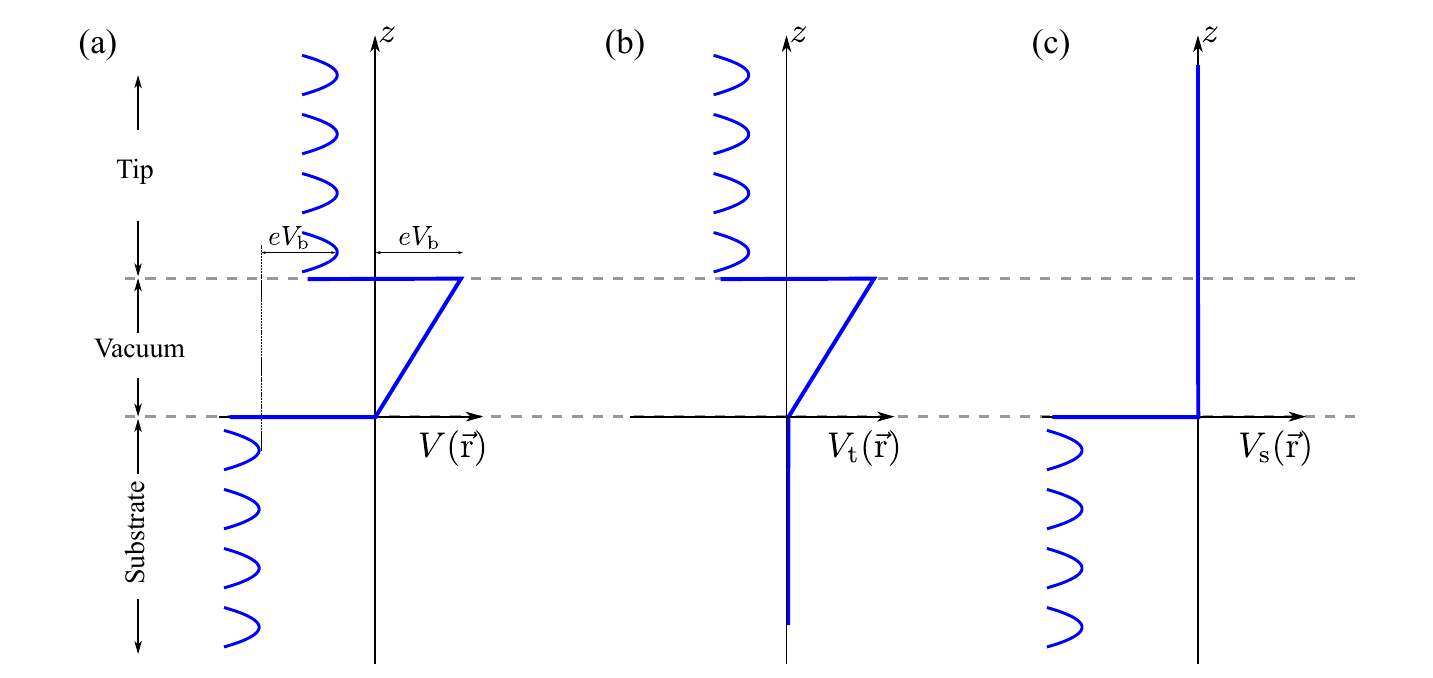}
\par\end{centering}
\caption{\label{fig:SM_potential}The schematic diagrams of the potential along
z-axis. (a) The potential $V\left(\vec{\mathbf{r}}\right)$ of the
tip and substrate. (b) The potential $V_{\mathrm{t}}\left(\vec{\mathbf{r}}\right)$
for the tip. (c) The potential $V_{\mathrm{s}}\left(\vec{\mathbf{r}}\right)$
for the substrate.}
\end{figure}

In this section, we show the details to the wave function of the tunneling
electron Hamiltonian,
\begin{equation}
H_{\mathrm{el}}=-\frac{1}{2m_{e}}\nabla^{2}+V\left(\vec{\mathbf{r}}\right).\label{eq:H_of_tunneling_electron}
\end{equation}
The total potential $V\left(\vec{\mathbf{r}}\right)$, illustrated
in Fig. \ref{fig:SM_potential}(a), is divided into two parts: the
tip $V_{\mathrm{t}}\left(\vec{\mathbf{r}}\right)$ (subfigure (b))
, and the substrate part $V_{\mathrm{s}}\left(\vec{\mathbf{r}}\right)$
(subfigure (c)). We use the approximate method proposed by Bardeen
in 1961 \citep{bardeen_1961,TUTORIAL_2006}. The Hamiltonian of the
free tip and substrate, $H_{\mathrm{el,t}}=-\nabla^{2}/2m_{e}+V_{\mathrm{t}}\left(\vec{\mathbf{r}}\right)$
and $H_{\mathrm{el,s}}=-\nabla^{2}/2m_{e}+V_{\mathrm{s}}\left(\vec{\mathbf{r}}\right)$.
For zero bias $V_{\mathrm{b}}=0$, the eigenstates of the free tip
and substrate are

\begin{subequations}
\begin{align}
H_{\mathrm{el,t}}|_{V_{\mathrm{b}}=0}\left|\phi_{k}\right\rangle  & =\xi_{k}\left|\phi_{k}\right\rangle ,\label{eq:solution_tip=000026base_zero_bias}\\
H_{\mathrm{el,s}}|_{V_{\mathrm{b}}=0}\left|\varphi_{n}\right\rangle  & =E_{n}\left|\varphi_{n}\right\rangle ,
\end{align}
\end{subequations}where $H_{\mathrm{el,t\left(s\right)}}|_{V_{\mathrm{b}}=0}$
represents the free tip (substrate) Hamiltonian at zero bias and $\left|\phi_{k}\right\rangle \left(\left|\varphi_{n}\right\rangle \right)$
is the eigenstate of free tip (substrate) with energy $\xi_{k}\left(E_{n}\right)$.
As the tip apex has been modeled as a metal sphere, its wave function
in the vacuum region has the asymptotic spherical form
\begin{equation}
\phi_{k}\left(\vec{\mathbf{r}}\right)=A_{k}\frac{e^{-\kappa_{k}\left|\vec{\mathbf{r}}-\vec{\mathbf{a}}\right|}}{\kappa_{k}\left|\vec{\mathbf{r}}-\vec{\mathbf{a}}\right|},
\end{equation}
where $\vec{\mathbf{a}}$ is the tip's center of curvature and $\kappa_{k}=\sqrt{-2m_{e}\xi_{k}}$
is its decay factor. $A_{k}$ can be determined by the first-principles
calculations. On the other hand, in the vacuum region, we take substrate's
wave function as
\begin{equation}
\varphi_{n}\left(\vec{\mathbf{r}}\equiv\left(x,y,z\right)\right)=B_{n}e^{-\kappa_{n}\left|z\right|},
\end{equation}
where $\kappa_{n}=\sqrt{-2m_{e}E_{n}}$ is the decay factor.

For nonzero bias $V_{\mathrm{b}}\neq0$, we take the potential change
induced by bias voltage as a perturbation and obtain the solution
up to the first-order correction,

\begin{subequations}
\begin{align}
H_{\mathrm{el,t}}\left|\phi_{k}\right\rangle  & \simeq\widetilde{\xi}_{k}\left|\phi_{k}\right\rangle ,\label{eq:solution_tip=000026base_nonzero_bias}\\
H_{\mathrm{el,s}}\left|\varphi_{n}\right\rangle  & \simeq\widetilde{E}_{n}\left|\varphi_{n}\right\rangle ,
\end{align}
\end{subequations}where $H_{\mathrm{el,t\left(s\right)}}$ represents
the free tip (substrate) Hamiltonian at bias $V_{\mathrm{b}}$ and
$\widetilde{\xi}_{k}\equiv\xi_{k}+eV_{\mathrm{b}}\left(\widetilde{E}_{n}\equiv E_{n}\right)$
is the corrected energy of state $\left|\phi_{k}\right\rangle \left(\left|\varphi_{n}\right\rangle \right)$.
Here we neglect the change to the wave function of the tip induced
by the applied voltage \citep{chen_1990}.

\section{The electron-molecule interaction}

In this section, we show the detailed derivation of the the effective
electron-dipole interaction between a tunneling electron and a single
molecule. The Coulomb interaction between a tunneling electron and
the molecule is written as
\begin{equation}
H_{\mathrm{el-m}}=\sum_{n=1}^{N}\left(-\frac{Z_{n}e^{2}}{\left|\vec{\mathbf{r}}-\vec{\mathbf{R}}_{n}\right|}+\frac{Z_{n}e^{2}}{\left|\vec{\mathbf{r}}-\vec{\mathbf{r}}_{n}\right|}\right),\label{eq:electron_molecule_interaction}
\end{equation}
where $\vec{\mathbf{r}}$ is the position of the tunneling electron.
The molecule contains $N$ bonds. For the $n$-th bond, $\vec{\mathbf{R}}_{n}\left(\vec{\mathbf{r}}_{n}\right)$
is the position of positive (negative) charge with effective charge
$Z_{n}$. $\vec{\mathbf{R}}_{0}\equiv\sum_{n=1}^{N}\vec{\mathbf{R}}_{n}Z_{n}/\sum_{n=1}^{N}Z_{n}$
denotes the center of the positive charge. For the case where the
distance between the tunneling electron and the molecule is much larger
than the size of the molecule, i.e., $\left|\vec{\mathbf{r}}-\vec{\mathbf{R}}_{0}\right|\gg\left|\vec{\mathbf{R}}_{n}-\vec{\mathbf{R}}_{0}\right|,\left|\vec{\mathbf{r}}_{n}-\vec{\mathbf{R}}_{0}\right|$
for all $n$, the coupling in Eq. (\ref{eq:electron_molecule_interaction})
becomes
\begin{align}
H_{\mathrm{el-m}} & =\sum_{n=1}^{N}\left(\frac{Z_{n}e^{2}}{\left|\vec{\mathbf{r}}-\vec{\mathbf{r}}_{n}\right|}-\frac{Z_{n}e^{2}}{\left|\vec{\mathbf{r}}-\vec{\mathbf{R}}_{n}\right|}\right)\nonumber \\
 & =\sum_{n=1}^{N}\left(\frac{Z_{n}e^{2}}{\left|\vec{\mathbf{r}}-\vec{\mathbf{R}}_{0}+\vec{\mathbf{R}}_{0}-\vec{\mathbf{r}}_{n}\right|}-\frac{Z_{n}e^{2}}{\left|\vec{\mathbf{r}}-\vec{\mathbf{R}}_{0}+\vec{\mathbf{R}}_{0}-\vec{\mathbf{R}}_{n}\right|}\right)\nonumber \\
 & =\sum_{n=1}^{N}\frac{Z_{n}e^{2}}{\left|\vec{\mathbf{r}}-\vec{\mathbf{R}}_{0}\right|}\left(\frac{1}{\left(1+\frac{2\left(\vec{\mathbf{r}}-\vec{\mathbf{R}}_{0}\right)\cdot\left(\vec{\mathbf{R}}_{0}-\vec{\mathbf{r}}_{n}\right)}{\left|\vec{\mathbf{r}}-\vec{\mathbf{R}}_{0}\right|^{2}}+\frac{\left|\vec{\mathbf{R}}_{0}-\vec{\mathbf{r}}_{n}\right|^{2}}{\left|\vec{\mathbf{r}}-\vec{\mathbf{R}}_{0}\right|^{2}}\right)^{1/2}}-\frac{1}{\left(1+\frac{2\left(\vec{\mathbf{r}}-\vec{\mathbf{R}}_{0}\right)\cdot\left(\vec{\mathbf{R}}_{0}-\vec{\mathbf{R}}_{n}\right)}{\left|\vec{\mathbf{r}}-\vec{\mathbf{R}}_{0}\right|^{2}}+\frac{\left|\vec{\mathbf{R}}_{0}-\vec{\mathbf{R}}_{n}\right|^{2}}{\left|\vec{\mathbf{r}}-\vec{\mathbf{R}}_{0}\right|^{2}}\right)^{1/2}}\right)\nonumber \\
 & \simeq\sum_{n=1}^{N}\frac{Z_{n}e^{2}}{\left|\vec{\mathbf{r}}-\vec{\mathbf{R}}_{0}\right|}\left(\left(1-\frac{\left(\vec{\mathbf{r}}-\vec{\mathbf{R}}_{0}\right)\cdot\left(\vec{\mathbf{R}}_{0}-\vec{\mathbf{r}}_{n}\right)}{\left|\vec{\mathbf{r}}-\vec{\mathbf{R}}_{0}\right|^{2}}\right)-\left(1-\frac{\left(\vec{\mathbf{r}}-\vec{\mathbf{R}}_{0}\right)\cdot\left(\vec{\mathbf{R}}_{0}-\vec{\mathbf{R}}_{n}\right)}{\left|\vec{\mathbf{r}}-\vec{\mathbf{R}}_{0}\right|^{2}}\right)\right)\nonumber \\
 & =\sum_{n=1}^{N}\frac{Z_{n}e^{2}}{\left|\vec{\mathbf{r}}-\vec{\mathbf{R}}_{0}\right|^{3}}\left(\vec{\mathbf{r}}-\vec{\mathbf{R}}_{0}\right)\cdot\left(\vec{\mathbf{r}}_{n}-\vec{\mathbf{R}}_{n}\right)\nonumber \\
 & =-e\frac{\left(\vec{\mathbf{r}}-\vec{\mathbf{R}}_{0}\right)\cdot\vec{\mathbf{\mu}}}{\left|\vec{\mathbf{r}}-\vec{\mathbf{R}}_{0}\right|^{3}},
\end{align}
where $\vec{\mathbf{\mu}}=\sum_{n=1}^{N}Z_{n}e\left(\vec{\mathbf{R}}_{n}-\vec{\mathbf{r}}_{n}\right)=-Ze\left(\vec{\mathbf{R}}_{0}-\vec{\mathbf{r}}_{0}\right),\;\left(Z\equiv\sum_{n=1}^{N}Z_{n},\vec{\mathbf{r}}_{0}\equiv\sum_{n=1}^{N}Z_{n}\vec{\mathbf{r}}_{n}/Z\right)$
denotes the total electric dipole moment of the molecule \citep{Minkin_Dipole_Moment_1970}.
We set the position of the positive charge as the origin of the coordinate
axes, i.e., $\vec{\mathbf{R}}_{0}=0$. The electron-molecule interaction
is expressed as
\begin{equation}
H_{\mathrm{el-m}}\simeq-e\frac{\vec{\mathbf{r}}\cdot\vec{\mathbf{\mu}}}{\left|\vec{\mathbf{r}}\right|^{3}}.
\end{equation}

\section{The photon count rate}

Once excited to its excited state, the molecule will decay to its
lower state and emits a photon spontaneously. Its photon counting
rate is proportional to its probability in excited state, $p_{e}\left(t\right)\equiv\sum_{n,k}\left|c_{e,k}\left(t\right)\right|^{2}$.
Taking the molecular excitation and the spontaneous emission process
together, we obtain the master equation for this excitation probability
\begin{align}
\frac{d}{dt}p_{e}\left(t\right) & =-\gamma p_{e}\left(t\right)+\frac{I_{\mathrm{inela}}}{e},
\end{align}
where $\gamma$ is the molecular spontaneous emission rate. In the
steady state, the excitation probability becomes $p_{e}\left(t\right)=I_{\mathrm{inela}}/e\gamma$.
Then the photon counting rate becomes
\begin{equation}
\Gamma=\gamma\frac{I_{\mathrm{inela}}}{e\gamma}=\frac{I_{\mathrm{inela}}}{e}.
\end{equation}

\section{The ratio of emission intensity}

\begin{figure}
\begin{centering}
\includegraphics[scale=0.3]{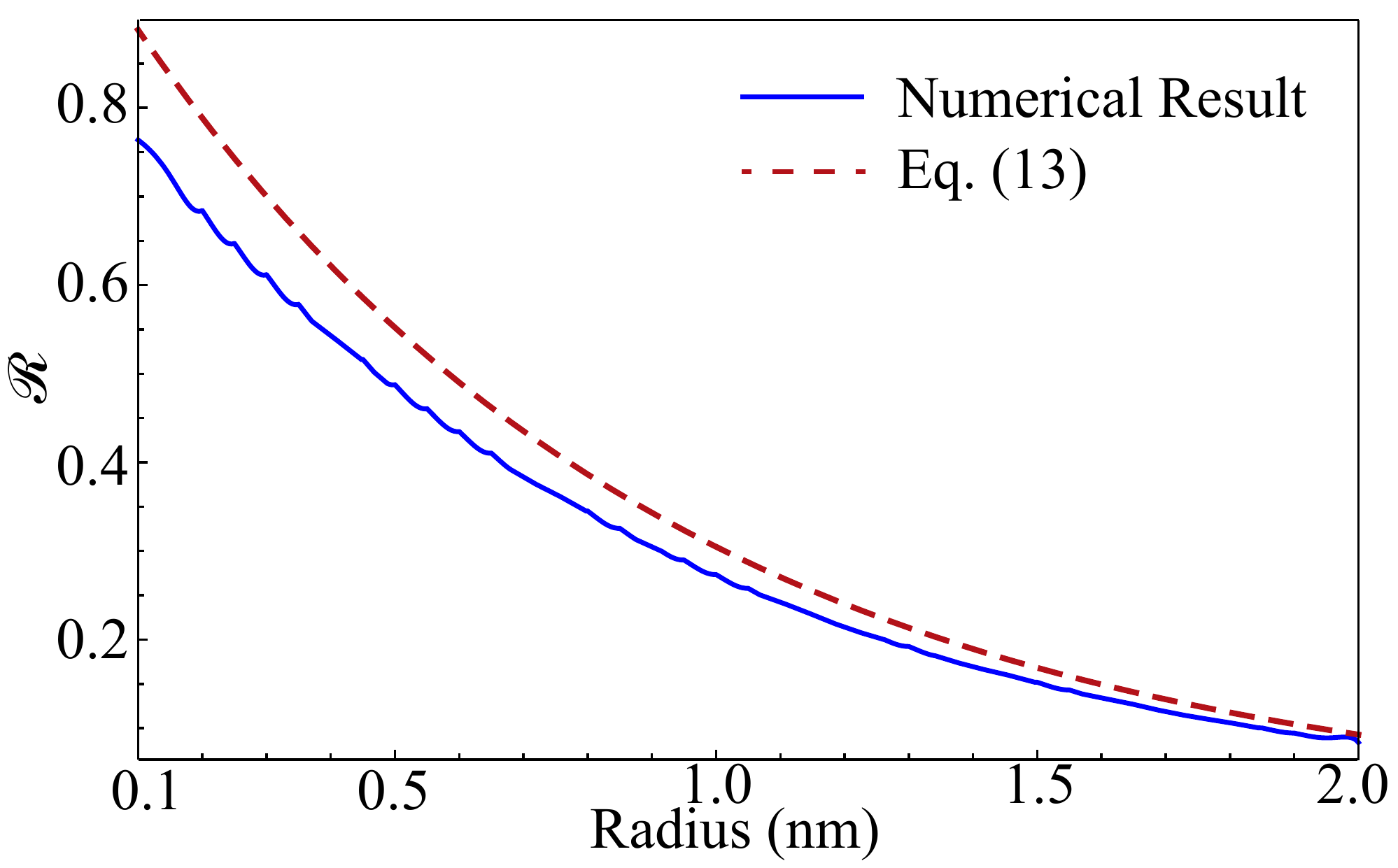}
\par\end{centering}
\caption{\label{fig:Photon-intensity-R_V} Photon intensity ratio versus the
radius of tip. The blue solid line represents the result obtained
by numerical calculation of the inelastic current and the red dashed
line shows the result given in Eq. (\ref{eq:I_2}). Here, the positive
bias is fixed to $+2.5$eV.}
\end{figure}

$\varphi_{n}\left(\vec{\mathbf{r}}\right)$ decays along the $+z$
direction \citep{Tersoff_Hamann_1983,Tersoff_Hamann_1985}. At positive
bias $V_{b}>0$, the transition matrix element of the inelastic tunneling
reads

\begin{align}
\mathscr{N}_{\mathrm{t,s}}|_{V_{\mathrm{b}},\xi_{k}\rightarrow E_{n}} & =\left\langle \phi_{k}\right|H_{\mathrm{e-m}}\left|\varphi_{n}\right\rangle \nonumber \\
 & =-e\mu_{z}\left\langle \phi_{k}\right|\frac{z}{\left|\vec{\mathbf{r}}\right|^{3}}\left|\varphi_{n}\right\rangle \nonumber \\
 & =-2\pi A_{k}B_{n}e\mu_{z}\int_{0}^{d}dz\int_{0}^{\infty}ldl\frac{e^{-\kappa_{k}\sqrt{x^{2}+y^{2}+\left(R+d-z\right)^{2}}}}{\kappa_{k}\sqrt{x^{2}+y^{2}+\left(R+d-z\right)^{2}}}\frac{z}{\sqrt{z^{2}+l^{2}}^{3}}e^{-\kappa_{n}z}\nonumber \\
 & \simeq-2\pi A_{k}B_{n}e\mu_{z}\int_{0}^{d}dz\int_{0}^{\infty}ldl\frac{e^{-\kappa_{k}\left(R+d-z\right)}}{\kappa_{k}\sqrt{x^{2}+y^{2}+\left(R+d-z\right)^{2}}}\frac{z}{\sqrt{z^{2}+l^{2}}^{3}}e^{-\kappa_{n}z}\nonumber \\
 & =-2\pi A_{k}B_{n}e\mu_{z}e^{-\left(\kappa_{k}-\kappa_{n}\right)R}\int_{0}^{d}dz\int_{0}^{\infty}ldl\frac{e^{-\kappa_{n}\left(R+d-z\right)}}{\kappa_{k}\sqrt{x^{2}+y^{2}+\left(R+z\right)^{2}}}\frac{d-z}{\sqrt{\left(d-z\right)^{2}+l^{2}}^{3}}e^{-\kappa_{k}z}\nonumber \\
 & \simeq e^{-\left(\kappa_{k}-\kappa_{n}\right)R}\mathscr{N}_{\mathrm{s,t}}|_{-V_{\mathrm{b}},\xi_{k}\rightarrow E_{n}},\label{eq:transition_element_relation}
\end{align}
where $\mathscr{N}_{\mathrm{t,s}}|_{V_{\mathrm{b}},\xi_{k}\rightarrow E_{n}}$
is the transition matrix element of the inelastic tunneling from tip's
state $\phi\left(\vec{\mathbf{r}}\right)$ with energy $\xi_{k}+eV_{\mathrm{b}}$
to substrate's state $\varphi\left(\vec{\mathbf{r}}\right)$ with
energy $E_{n}$ at bias $V_{\mathrm{b}}$ and $\mathscr{N}_{\mathrm{s,t}}|_{-V_{\mathrm{b}},\xi_{k}\rightarrow E_{n}}$
is the transition matrix element of the inelastic tunneling from substrate's
state $\varphi\left(\vec{\mathbf{r}}\right)$ with energy $\xi_{k}$
to tip's state $\phi\left(\vec{\mathbf{r}}\right)$ with energy $E_{n}-e\left|V_{\mathrm{b}}\right|$.
Without loss of generality, we have chosen the tip right above the
molecule ($\left|OA\right|=0$) and the molecular dipole in the $z$
direction ($\mu_{z}\neq0$ while $\mu_{x}=\mu_{y}=0$). In deriving
Eq. (\ref{eq:transition_element_relation}), we have taken the approximation
that, in the vacuum region, tip's state only decays in $z$ direction.
We see that for the same initial energy, the ratio of the transition
matrix element at positive bias $V_{\mathrm{b}}$ to that at negative
bias $-V_{\mathrm{b}}$ is $e^{-\left(\kappa_{k}-\kappa_{n}\right)R}$.
Then, the inelastic tunneling current at positive bias is
\begin{align}
I_{+\mathrm{,inela}}|_{V_{\mathrm{b}}} & \simeq2\pi e\int_{\mu_{0}}^{\mu_{0}+eV_{\mathrm{b}}-E_{eg}}dE_{n}\rho_{\mathrm{s}}\left(E_{n}\right)\rho_{\mathrm{t}}\left(\xi_{k}\right)\left|\mathscr{N}_{\mathrm{t,s}}|_{V_{\mathrm{b}},\xi_{k}\rightarrow E_{n}}\right|^{2}|_{\xi_{k}=E_{n}-eV_{\mathrm{b}}+E_{eg}}\mathbb{R}\nonumber \\
 & \simeq2\pi e\int_{\mu_{0}}^{\mu_{0}+eV_{\mathrm{b}}-E_{eg}}dE_{n}e^{-2\sqrt{-2m_{e}E_{n}}\left(\sqrt{1-\frac{eV_{\mathrm{b}}-E_{eg}}{E_{n}}}-1\right)R}\rho_{\mathrm{s}}\left(E_{n}\right)\rho_{\mathrm{t}}\left(\xi_{k}\right)\left|\mathscr{N}_{\mathrm{s,t}}|_{-V_{\mathrm{b}},\xi_{k}\rightarrow E_{n}}\right|^{2}|_{\xi_{k}=E_{n}-eV_{\mathrm{b}}+E_{eg}},\label{eq:I_2_1}
\end{align}
where $\rho_{\mathrm{t}}\left(E\right)\,\left(\rho_{\mathrm{s}}\left(E\right)\right)$
denotes the density of state of tip (substrate) at the energy $E$
and $\mu_{\mathrm{0}}$ is the Fermi energy of substrate and tip at
zero bias. For a small bias which satisfies $\left(eV_{\mathrm{b}}-E_{eg}\right)/\mu_{0}\ll1$,
Eq. (\ref{eq:I_2_1}) can be further simplified as
\begin{align}
I_{+,\mathrm{inela}}|_{V_{\mathrm{b}}} & \simeq2\pi e\int_{\mu_{0}}^{\mu_{0}+eV_{\mathrm{b}}-E_{eg}}dE_{n}e^{\sqrt{-2m_{e}E_{n}}\left(\frac{eV_{\mathrm{b}}-E_{eg}}{E_{n}}\right)R}\rho_{\mathrm{s}}\left(E_{n}\right)\rho_{\mathrm{t}}\left(\xi_{k}\right)\left|\mathscr{N}_{\mathrm{s,t}}|_{-V_{\mathrm{b}},\xi_{k}\rightarrow E_{n}}\right|^{2}|_{\xi_{k}=E_{n}-e\mathrm{V_{b}}+E_{eg}}\nonumber \\
 & \simeq e^{-R\sqrt{-2m_{e}\mu_{0}}\left(\frac{eV_{\mathrm{b}}-E_{eg}}{-\mu_{0}}\right)}2\pi e\rho_{\mathrm{s}}\left(\mu_{0}\right)\rho_{\mathrm{t}}\left(\mu_{0}\right)\int_{\mu_{0}}^{\mu_{0}+e\mathrm{V_{b}}-E_{eg}}dE_{n}\left|\mathscr{N}_{\mathrm{s,t}}|_{-V_{\mathrm{b}},\xi_{k}\rightarrow E_{n}}\right|^{2}|_{\xi_{k}=E_{n}-e\mathrm{V_{b}}+E_{eg}}\nonumber \\
 & \simeq e^{-R\sqrt{-2m_{e}\mu_{0}}\left(\frac{eV_{\mathrm{b}}-E_{eg}}{-\mu_{0}}\right)}I_{-\mathrm{,inela}}|_{-V_{\mathrm{b}}}.\label{eq:I_2}
\end{align}
For noble metal, such as Au, Ag, and Cu, its density of state is almost
a constant for several electron volt around its Fermi energy \citep{handbook}.
Thus, in deriving Eq. (\ref{eq:I_2}), we have extracted the density
of state function from the integral. Finally, we see in Eq. (\ref{eq:I_2})
that this ratio decays with tip's radius $R$ and bias voltage $V_{\mathrm{b}}$.

Fig. \ref{fig:Photon-intensity-R_V} shows the ratio $I_{+\mathrm{,inela}}|_{V_{\mathrm{b}}}/I_{-\mathrm{,inela}}|_{-V_{\mathrm{b}}}$
with bias and tip's radius of curvature. The blue line represents
the numerical result and the red line shows the analytical result
presented in Eq.(15) of the main context. Fig. \ref{fig:Photon-intensity-R_V}
shows that the ratio decays with the increase of tip's radius. And
the larger the radius is, the better the two results coincide with
each other. The underlying reason is that the main contribution of
the integral in the transition matrix element comes from a small region
between the tip and the molecule. And in deriving Eq. (\ref{eq:I_2}),
we also have taken the approximation that, in the vacuum, tip's wave
only decays in $z$ direction. Thus, for a tip with a larger radius,
tip's wave function behaves more like a wave that decays only in $z$
direction in the small region under tip.

\bibliographystyle{apsrev4-1}